\documentclass{aastex}

\begin{document}

\title{In-orbit Calibrations of the Ultra-Violet Imaging Telescope}
\author{S.N. Tandon$^{1,2}$, Annapurni Subramaniam$^{2}$, 
V. Girish$^{3}$,
J. Postma$^{4}$,
K. Sankarasubramanian$^{2,3,8}$,
S. Sriram${^2}$,
C.S. Stalin$^{2}$,
C. Mondal$^{2}$, 
S. Sahu$^{2}$,
P. Joseph${^2}$,
J. Hutchings$^{5}$,
S.K. Ghosh$^{6,7}$,
I.V. Barve${^2}$,
K. George$^{2}$,
P.U. Kamath$^{2}$,
S. Kathiravan$^{2}$,
A. Kumar$^{2}$,
J.P. Lancelot$^{2}$,
D. Leahy$^{4}$,
P.K. Mahesh$^{2}$,
R. Mohan$^{2}$,
S. Nagabhushana$^{2}$,
A.K. Pati$^{2}$,
N. Kameswara Rao$^{2}$, 
Y.H. Sreedhar$^{2}$ \and
P. Sreekumar$^{2}$ 
}
\affil{
$^1$Inter-University Center for Astronomy and Astrophysics, Pune, India\\
$^2$Indian Institute of Astrophysics, Koramangala II Block, Bangalore-560034, India\\
$^3$ISRO Satellite Centre, HAL Airport Road, Bangalore 560017\\
$^4$University of Calgary, 2500 University Drive NW, Calgary, Alberta Canada\\
$^5$National Research Council of Canada, Herzberg Astronomy and Astrophysics, 5071 West Saanich Road, Victoria, BC V9E 2E7, Canada\\
$^6$National Centre for Radio Astrophysics, Pune, India\\
$^7$Tata Institute of Fundamental Research, Mumbai, India\\
$^8$Center of Excellence in Space Sciences India, Indian Institute of Science Education and Research (IISER), Kolkata,
Mohanpur 741246, West Bengal, India}
\email{purni@iiap.res.in}


\begin{abstract}
The Ultra-Violet Imaging Telescope (UVIT) is one of the payloads in ASTROSAT, the first Indian Space Observatory.
The UVIT instrument has two 375~mm telescopes: one for the far-ultraviolet (FUV) channel (1300--1800~\AA), 
and the other for the near-ultraviolet (NUV) channel (2000--3000~\AA) and the visible (VIS) channel (3200--5500~\AA). UVIT is primarily designed for simultaneous
imaging in the two ultraviolet channels with spatial resolution better than 1.8\arcsec, along with provision for slit-less spectroscopy in the NUV and FUV channels.The results of in-orbit calibrations of UVIT are presented in this paper.
\end{abstract}


\section{Introduction} \label{sec:intro}

The Ultra-Violet Imaging Telescope (UVIT) is the ultra-violet eye of the multi-wavelength satellite ASTROSAT, 
launched on September 28, 2015 by the Indian Space Research Organisation (ISRO). UVIT is designed to make images, 
simultaneously in NUV (2000-3000~\AA) and FUV (1300-1800~\AA) wavelengths, in a field of $\sim$ 28\arcmin, with a FWHM $<$1.8\arcsec. 
The sensitivity in the FUV is $\sim$ 20 mag in the AB magnitude scale for a  200s exposure. 
Low resolution slit-less spectroscopy can also be done in the NUV and FUV. In this paper we describe details of the
in-orbit calibrations, as well as report the results of ground calibrations which complement the in-orbit calibrations.  Some of the initial results
of calibration as well as some early science results can be found in Tandon et al. (2017) and Subramaniam et al. (2016a). Here, we present the results of full calibration for all the filters and gratings in FUV and NUV channels.

 The UVIT has three times better spatial resolution when compared to GALEX and has multiple filter in the FUV and NUV channels. A comparison of the UVIT with respect to other UV missions is given in Tandon et al. (2017).
 Other missions which have common features with UVIT are Galex (Martin et al. 2005 and Morrissey et al. 2007), Swift-UVOT (Roming et al. 2005) and XMM-OM (Mason et al. 2001) . We compare the features of UVIT with these  in Table 1. A similar description is also presented in Tandon et al. (2017).
 \begin{table*}[h]
\begin{center}
\caption{Comparison of UVIT with other similar missions are tabulated below }
\begin{tabular}{lcccc}
\hline
Parameter & UVIT &GALEX& SWIFT-UVOT & XMM-OM\\
\hline
Pass bands & FUV, NUV& FUV, NUV & NUV, VIS & NUV, VIS \\
Filters within a band& YES & NO & YES & YES \\
Slit-less spectroscopy& YES & YES & YES & YES \\
Field of View (diameter) & 28\arcmin & 1.$^o$2 & 17\arcmin$\times$17\arcmin & 17\arcmin$\times$17\arcmin \\
Effective Area (NUV) & $\sim$ 50 & similar& similar & similar\\
Effective Area (FUV) & $\sim$ 10& Twice& --& --\\
Spatial Resolution (FWHM)(\arcsec) & $<$ 1.8\arcsec & 5\arcsec& $<$2\arcsec &$<$2\arcsec \\
Simultaneous in FUV \& NUV & YES & YES & NO & NO \\
\hline
\end{tabular}
\end{center}
\end{table*}

The UVIT has been operational for the last 19 months, where the first 4 months were dedicated for performance verification and
in-orbit calibrations, followed by proposal based observations for about one year. The performance of the telescope is also monitored using
regular sensitivity checks. The science targets observed during the period include star clusters, galaxies, galaxy clusters, AGN, Chandra deep field, exo-planets, planetary nebulae, supernovae remnants etc. Subramaniam et al. (2016b) presented the first science result from UVIT, based on initial calibrations. The calibrations presented here are essential to calibrate and derive science from all observations carried out using the UVIT.  Images obtained by the UVIT can be found in http://uvit.iiap.res.in/. Indian as well as  international astronomers have access to the observing time of ASTROSAT and UVIT through proposals.

The paper is arranged  as follows: Section 2 gives a brief description of the instrument and observations with it, Section 3 describes in details the procedures of calibrations and the results, Section 4 describes the calibrations to be done in the future, and Section 5 gives a summary.

\section{Instrument and Observations}

Details of the instrument and its function can be found in Tandon et al. (2017), Subramaniam et al. (2016a) and Kumar et al. (2012).
We give below a brief description.

\subsection{Instrument}
The instrument is configured as two co-aligned telescopes. Each telescope consists of f/12 Ritchey-Chretien optics, of aperture 375 mm, with filters and detectors. One telescope observes in FUV (1300-1800~\AA), and the other in  NUV (2000-3000~\AA) and VIS (3200-5500~\AA); the observations in VIS are primarily used for tracking aspect of the telescopes. For each of the three channels a filter wheel is used to select a filter, or a grating in FUV (1300-1800~\AA) and NUV (2000-3000~\AA). Each of the channels has an intensified CMOS imager  of aperture 39mm, which can work  either in photon-counting mode (with high intensification)  or in integration mode (with low intensification).  For each detector, a suitable window is chosen and appropriate photo-cathode is deposited on it. In all other details, the three detectors are identical.  Each detector has an aperture of 39 mm and the intensified image is reduced to $\sim$ 12.5 mm by a fibre-taper to match size of the imager. In the photon counting mode, photons are detected, by the hardware, in each frame of the imager as pixels of local maxima, within a window, which is above a threshold; the window can be selected as 3$\times$3 or 5$\times$5 pixels, but in the orbit only 3$\times$3 has been used so far. For each detected photon centroid is calculated for the signals in the window and sent as data; the centroids have systematic errors which depend on location of the event within the pixel and are corrected for in the analysis on ground. In the integration mode full images are sent as data. As readout for the images is done in rolling shutter mode, the average time of exposure is not identical for each row of the image and can be corrected for in specific observation if required.  The reader is referred to Tandon et al. (2017), Kumar et al. (2012)   and Postma et al. (2011), and the references therein for more details of the instrument.

\subsection{Observations}
The observations can be made for the full field of $\sim$ 28\arcmin~diameter or for partial field. The maximum frame-rate for the 
full field is $\sim$ 29 frames/s.  For partial fields the rate is higher and  a 
field of $5.5\times5.5$ arcmin$^2$ is read at a rate of $\sim$ 600 frames/s. The detectors for FUV and NUV are operated in photon counting mode, while the detector for VIS is operated in integration mode. 
To avoid saturation in FUV and NUV detectors, rate of photons for any point source (or any  3$\times$3 pixels of CMOS imager) should be $<< $ 1 photon/frame. 
During observations of any source, aspect of the satellite drifts up to $\sim$ 1\arcmin ~at a typical rate 
$\sim$ 1\arcsec/s which 
can increase several fold during the perturbations caused by rotation of SSM (an all sky X-ray monitor on ASTROSAT). Therefore, 
 FUV \& NUV images are recorded with exposure $<$ 35ms and stacked on ground with shift and add algorithm; the shift is found by the images 
taken in VIS (3200-5500~\AA),  at rate $\sim$ 16 frames/s, and stacking every 16 of these on-board to get an image for transmission to ground. Given these short exposures, any effects of readout in rolling shutter mode are not significant. Most of the observations are made with FUV and NUV detectors working in photon counting mode with full field, and the VIS detector working in integration mode. For specific observations, e.g. to get a higher rate of frames for relatively bright ultraviolet objects, the detectors are used for partial field. 
The key performance parameters of the three channels are tabulated in Table~2,
and properties of the filters available in the three channels are shown in Table~3. The effective areas of the filters as a function of wavelength
are shown in Figure 1, which are based on ground calibrations.
\begin{figure*}[h]
\begin{center}
\includegraphics[scale=0.2]{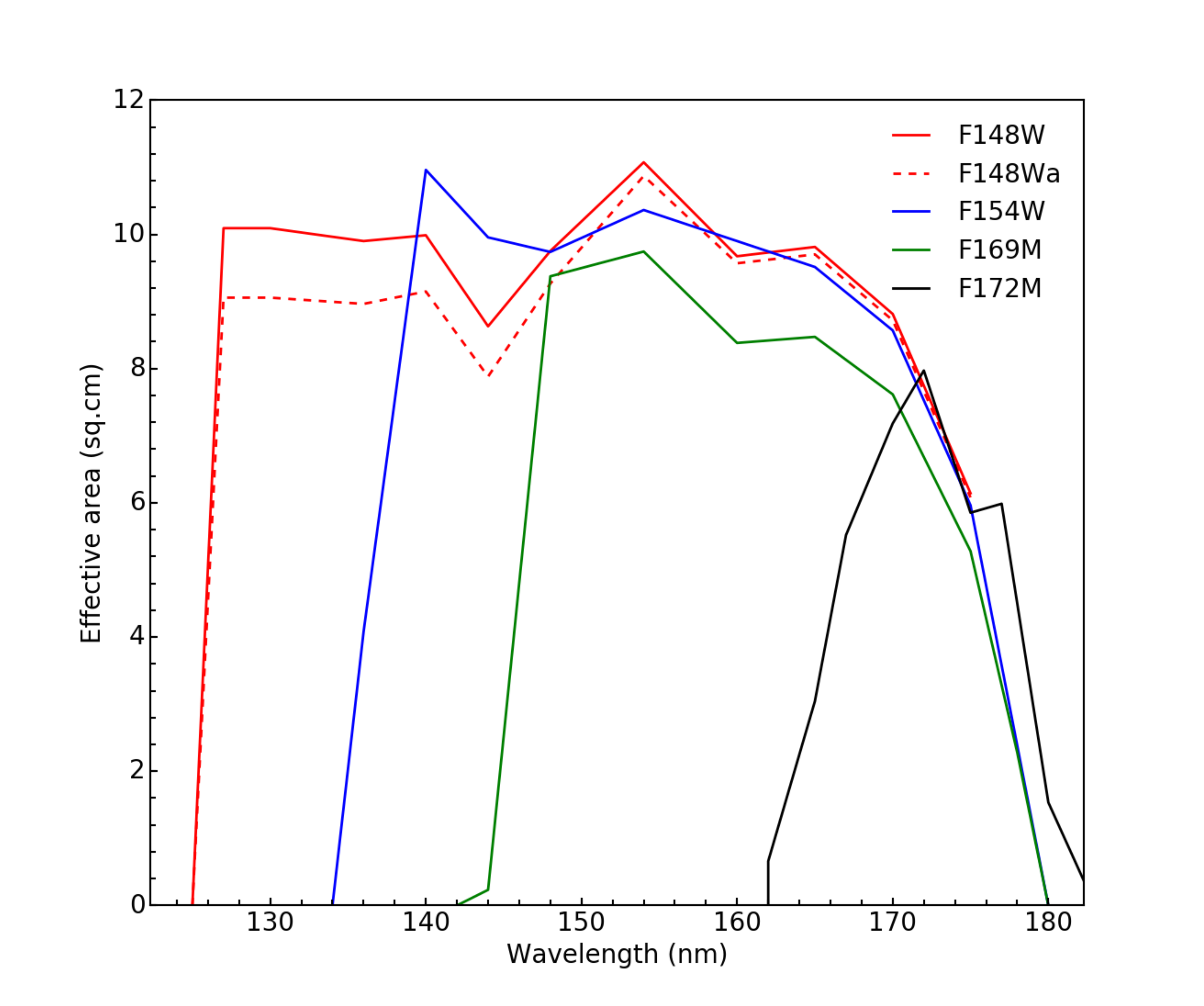}
\includegraphics[scale=0.2]{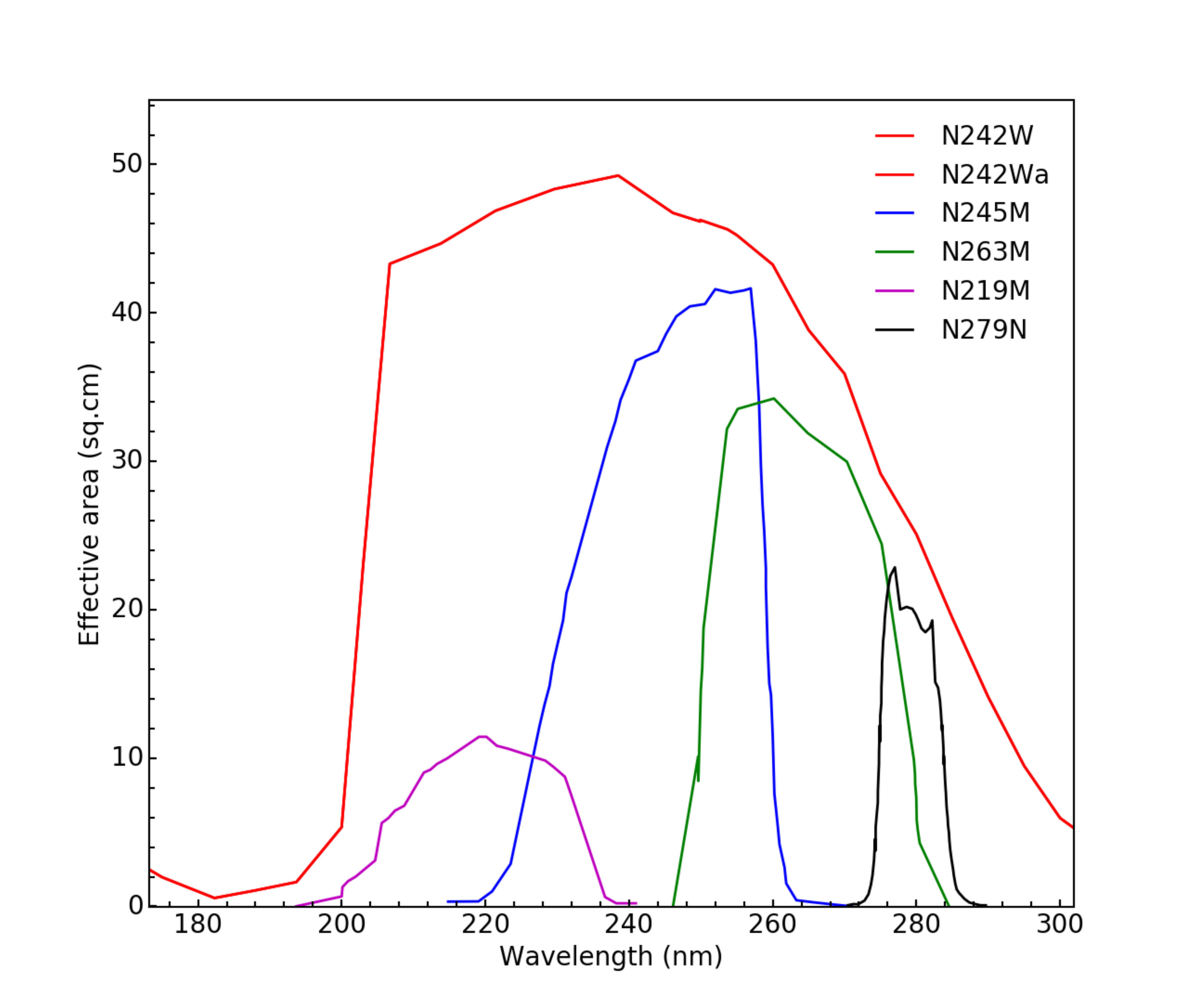}
\includegraphics[scale=0.2]{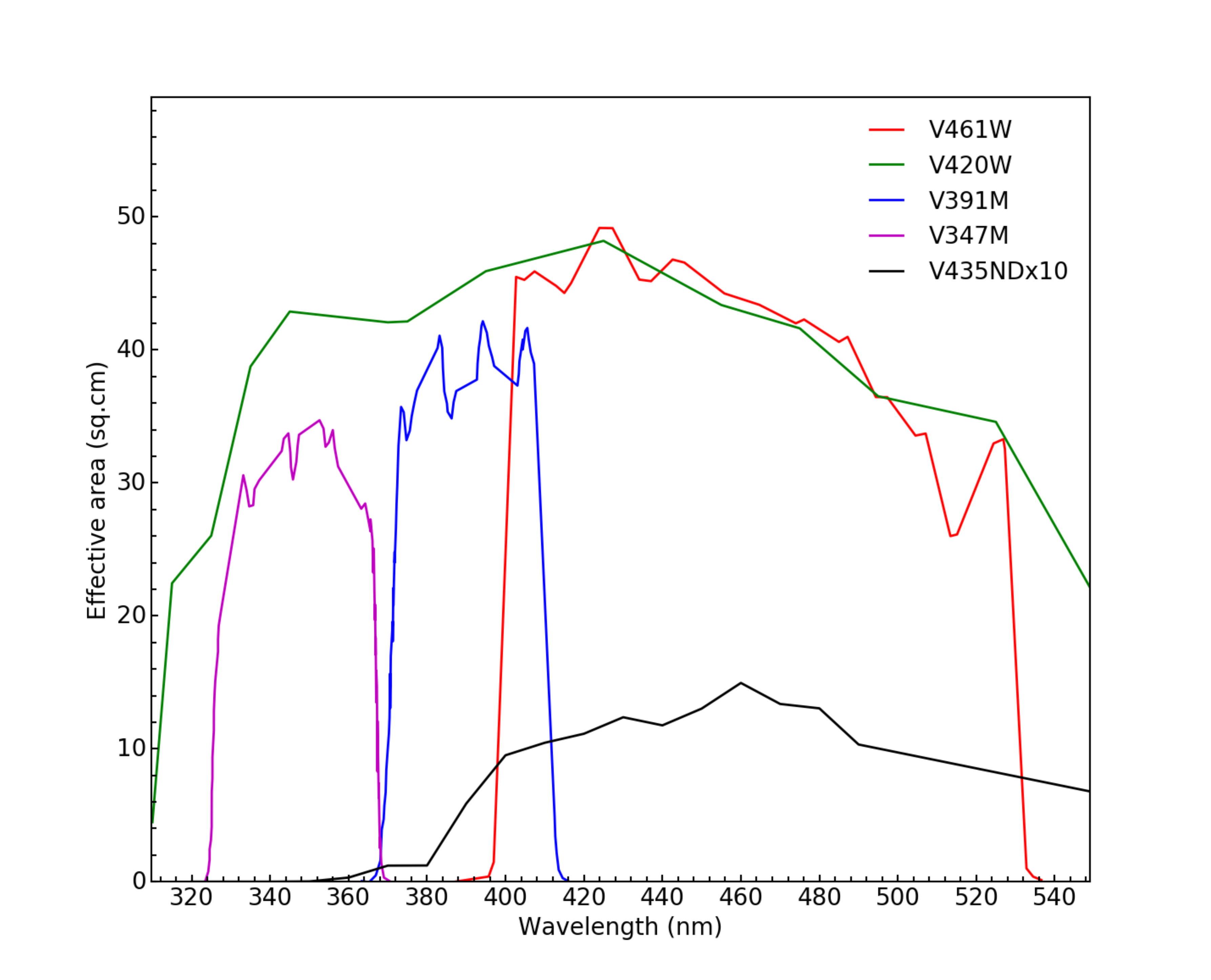}
\caption{The effective area curves are shown for the filters in FUV, NUV and VIS channels of the UVIT. These are based on ground calibrations.
} 
\end{center}
\end{figure*}

\begin{table*}[h]
\begin{center}
\caption{Performance parameters for the three channels are shown.$^a$ }
\begin{tabular}{lccc}
\hline
Parameter & FUV &NUV& VIS$^a$\\
\hline
Wavelength (\AA) & 1300-1800& 2000-3000 & 3200-5500 \\
Mean Wavelength$^b$ (\AA) & 1481 & 2418 & 4200 \\
Mean Effective Area (cm$^2$) & $\sim$ 10 & $\sim$ 40 & $\sim$ 50 \\
Field of View (diameter-arcmin) & 28 & 28 & 28 \\
Plate Scale (\arcsec/pixel) & 3.33 & 3.33 & 3.30 \\
Astrometric Accuracy (\arcsec)(rms) & 0.8 & 0.8 &—\\
Zero-point Magnitude$^c$ & 18.0 & 19.8 & — \\
Spatial Resolution$^d$ (FWHM)(\arcsec) &1.3 to 1.5 & 1.0 to 1.4& 2.5 \\
Spectral Resolution$^e$ (\AA) & 17  & 33 & \\
Saturation (counts/sec)$^f$ (10\%) & 6 & 6 \\
\hline
\end{tabular}
\end{center}
\vskip 1.0ex
\small{
{\bf Notes:} $^{a}$ For the VIS channel all the parameters are based on ground calibrations. This channel is operated in integration mode.
Photometry calibration is not done as we don't expect doing science with VIS channel observations. This channel is purely meant for aspect
calculation. \\
$^{b}$ The Mean Wavelength is obtained by weighing wavelength, for the filter with maximum bandwidth, with the effective area (for the filter with maximum bandwidth) as estimated by calibrations on the ground.\\
$^{c}$ The zero point magnitude (for the filter with maximum bandwidth) is in AB system and refers to the flux of HZ4 at the Mean Wavelength \\
$^{d}$ It depends on perturbations in the pointing.\\
$^{e}$ These are for the gratings.\\
$^{f}$ The saturation is given for the full field images. These are taken at a rate 28.7 frames/s; images for partial field are taken at higher frequency of the frames and the range of linearity is higher (see  section 3.1.2)\\}
\end{table*}

\begin{table*}[h]
\begin{center}
\caption{Properties of individual filters are shown for the three channels. Where $\lambda_{mean}$ is the Mean Wavelength and $\Delta \lambda$ is the Band Width as defined in Subsection 3.1.1}
\begin{tabular}{llcc}
\small
Filter Name & Filter& $\lambda_{mean}$ (\AA) & $\Delta \lambda$ (\AA) \\
\hline
FUV:&&&\\
\hline
F148W&CaF2-1&1481&500\\
F148Wa&CaF2-2&1485&500\\
F154W&BaF2&1541&380\\
F172M&Silica&1717&125\\
F169M&Sapphire&1608&290\\
\hline
NUV:&&&\\
\hline
N242W&Silica-1&2418&785\\
N242Wa&	Silica-2&2418&785\\
N245M&NUVB13&2447&280 \\
N263M&NUVB4&2632&275 \\
N219M&NUVB15&2196&270 \\
N279N&NUVN2&2792&90\\
\hline
VIS:&&&\\
\hline
V347M&VIS1&3466&400 \\
V391M&VIS2&3909&400 \\
V461W&VIS3&4614&1300 \\
V420W&BK7&4200&2200\\
V435ND&ND1&4354&2200\\
\hline
\end{tabular}
\end{center}
\vskip 1.0ex
\end{table*}

\begin{figure*}[h]
\begin{center}
\includegraphics[scale=0.7]{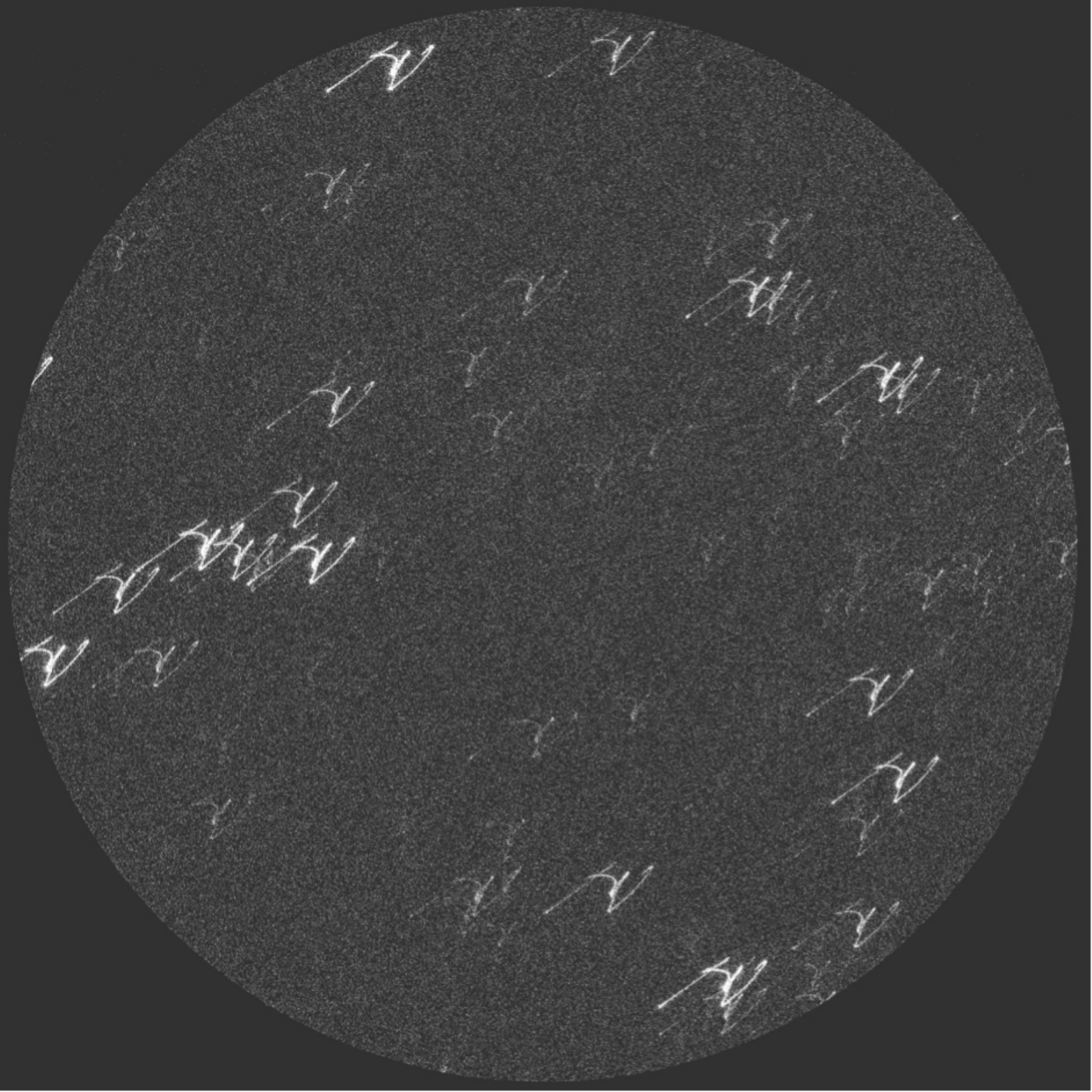}
\caption{A raw image of Abell 2256 is shown. Centroids for all the detected photons are used without any corrections. Each bright trail corresponds to a bright star. Shape of the trails is defined by drift of the S/C during the observation.
} 
\end{center}
\end{figure*}

\subsection{Backgrounds}
 In order to minimise the effect of scattered solar radiation and the radiation from bright side of the earth, observations are only made in dark side of the orbit. Depending on the level of solar-activity, in FUV geo-coronal lines of OI can give large background for observations with F148W filter, but with F154W filter most of it is eliminated. However, during the first six months after launch observations with F148W did not give large background. Zodiacal light makes a major contribution to the background in NUV. In addition to these backgrounds there is additional background from the Galaxy which depends on the Galactic latitude. The dark counts of the detector are negligible, but cosmic-ray interactions contribute $\sim$150 counts/sec in FUV and NUV, irrespective of the filter, for the full field.
 These interactions are seen in $\sim$ 3 frames every second as showers, each with an average of $\sim$ 50 events. The average number of events per shower is consistent with these being due to Cerenkov radiation of individual particles in the filters and windows of the detectors (see Viehmann \& Eubanks 1976;  Nasa Technical Note TN D-8147).  The expected number of primary cosmic rays contributing to such showers is about 1/s, and most of the showers seem to be due to  secondary particles produced during interactions in the satellite.
 As the average number of events is $\sim$50 per shower, it is possible to reject the corresponding frames by rejecting all the frames with counts beyond a threshold. This rejection leads to a loss of 10\% data if imaging is done for full field, and should only be applied for dark fields where the other backgrounds are not much more than 150 counts/sec. The overall background in NUV and FUV have a range,  24 - 25 mag per 10 arcsec$^2$ and 24 - 26 mag per 10 arcsec$^2$ respectively.     

\subsection{Analysis of Images}
The images from NUV and FUV detectors are received as a list of centroids (calculated to 1/32 of a pixel of the 512$\times$512 CMOS imager) of the detected photons in each frame. For VIS detector, signals for all pixels of the CMOS imager are received. As pointing of the S/C drifts by up to $\sim$1\arcmin, a raw image (obtained by plotting centroids of all the detected photons) would look like what is shown in Figure 2.  To make final image the following corrections are done for each individual centroid:
\begin{enumerate}
 \item bias in the coordinates due to the algorithm for centroiding (see Postma et al 2011), 
 \item shift in position corresponding to distortions in the detector  and the optics (see Section 3.4), and 
 \item effective number of photons as corrected for flat-field variations, as compared to centre of the field, as obtained in ground calibrations (see Section 3.1). In some frames two detected photons could fall within the window of 3$\times$3 pixel used to detect photons. In such cases, the two photons are detected as one, and the returned centroid corresponds to a weighted mean of positions for the two photons. Such double events would lead to underestimate of the flux and narrowing of the PSF.
 \item  shift in position corresponding to estimated shift of the frame (or the corresponding time) as compared to an arbitrarily chosen reference frame (or time), 
 \end{enumerate}
 
\section{In-orbit Calibrations}
In this section details of the in-orbit calibrations are described. The details are supplemented by a brief description of the relevant 
ground calibrations if needed.
\subsection{Photometric Calibrations}
The primary Photometric calibration for FUV and NUV channels was achieved by observing standard stars for which flux calibrated 
spectra are available in the UV wavelength range. The deliverables of the primary photometric calibration are the 
Zero Point magnitude  and the Unit Conversion factor for all the filters. The magnitude system adopted for the UVIT 
filters in the AB magnitude system (Oke 1974) and hence the magnitudes derived will be in this system. The Unit 
Conversion factor relates the flux of the source, at the Mean wavelength of a filter, to the observed count-rate. 
The calibrations of the VIS channel filters are not done as the images obtained are only used to find positions of the stars.

The primary photometric standard star should have flux calibrated spectrum available in the wavelength range 
covering the FUV and NUV filters of the UVIT. Such sources are available in the CALSPEC database of the HST and 
these are the potential targets for the primary photometric calibration. As these sources have flux calibrated 
spectra, we can predict the expected count rates based on the ground calibrations. In order to choose an appropriate standard
star for photometric calibration, a set of criteria was developed in order to achieve best possible calibration.
The following set of criteria was used to select an optimal standard star for photometric calibration.
\begin{itemize}
\item the star should be bright enough to obtain enough photons in a reasonable exposure time (1000 to 3000 s per filter), 
\item in all the filters, the count rate should not exceed the limit beyond which the saturation correction fails,
\item to detect variations in the sensitivity across the field, the same star is observed at various locations in the field, 
which requires that there is no bright source in the sky within a radius of 35' around the star with a flux that can damage the detector,  
\item should have a declination beyond the range $-$6 degrees to $+$6 degrees (a mission-constraint requires that the source be at 
least 12 degrees away from tangent to the orbit, which is inclined by 6 degrees to the equator, during the observation).
\end{itemize}
The standard source for calibration was adopted as HZ4, which is a moderately bright WD, and satisfies all the above criteria.
From the ground calibrations, we calculated the mean Effective Area, Mean Wavelength and Band Width of the filters.  The first 
estimates of the Zero Point magnitude and the Unit Conversion factor were based on the ground calibrations (shown in Table~4).
A comparison of the estimated count rates for the standard star with the observed count rates give corrections for the mean Effective
 Area, Zero Point magnitude, and Unit Conversion factor for the various filters. 

The equations which are used to derive the flux and magnitude of the observed object are:\\
\begin{equation}
 Flux (ergs~cm^{-2}~s^{-1}~\AA^{-1}) =  CPS \times Unit~Conversion 
\end{equation}
\begin{equation}
 Magnitude (AB system) = -2.5~log~(CPS) + Zero~Point,
\end{equation}
where CPS corresponds to observed,  background-subtracted counts per sec with a filter, and Unit Conversion factor and Zero Point magnitude are for the filter. 
The above equations are similar to those derived for the GALEX filters (Morrissey et al. 2007).

\subsubsection{Definitions}
We present the definitions of the parameters used for the photometric calibration, such as, Mean Wavelength, Band Width, 
Calculated-mean Effective Area (CEA), Estimated-mean Effective Area (EEA), Unit Conversion factor (UC), and Zero Point magnitude (ZP).  Of these parameters, the final values of the first three are obtained in ground calibrations while those of the last three are derived from in-orbit calibrations.
\itemize
\item Mean Wavelength ($\lambda_{mean}$) is the mean of wavelengths weighted with effective area. It is given by Eq. 3
\begin{equation}
\lambda_{mean}  = \frac{\int \lambda EA(\lambda) d\lambda)}{\int EA(\lambda) d\lambda)}                     
\end{equation}
where EA($\lambda$)  is the effective area (in $cm^{2}$) at wavelength $\lambda$ (in \AA) measured in the ground calibrations.
\item Band Width is defined by the wavelengths where effective area falls to 50\% of its peak value, 
as obtained in the ground calibrations.
\item Calculated mean Effective Area (CEA) is the mean of EA($\lambda$)  between the wavelengths where effective 
area falls to 50\% of its peak value. 
\begin{equation}
CEA = \frac {\int EA(\lambda) d\lambda)} {\int d\lambda}
\end{equation}
where the integration limits are the wavelengths where effective area is 50\% of its peak value.                                                               
\item Unit Conversion factor is the flux at Mean Wavelength which gives one detected photon per second (1 CPS). 
Its definition is tied to the spectrum of calibration source (HZ4). The flux for the source at Mean Wavelength 
is defined as an average over the band by the following equation:
\begin{equation}
F(\lambda_{mean})  = \frac{\int F(\lambda) EA(\lambda) d\lambda}{(CEA \times Band~Width)}
\end{equation}
where  $F(\lambda)$ is the standard flux of the source at $\lambda$.
Unit Conversion can be written as 
\begin{equation}
Unit~Conversion = \frac{F(\lambda_{mean})}{(CPS)} 
\end{equation}
As the right hand side of Eq. 5 involves a division of EA($\lambda$) by CEA, it can be seen that the measurement of {\it Unit Conversion} 
only depends on the relative values of EA($\lambda$) at different wavelengths. The measured value of 
Estimated-mean Effective Area (EEA) can be written as
\begin{equation}
EEA = \frac {(Measured~CPS)}{ (Band~width \times F(\lambda_{mean}))}
\end{equation}
\item The UVIT magnitudes are in AB magnitude system, based on the following definition:\\
\begin{equation}
m_{AB} = -2.5log_{10}f_{\nu}-48.6
\end{equation}
where the units of $f_{\nu} = f_{\lambda}*(\lambda^{2}/c)$ are CGS.
Zero Point magnitude is defined as the AB magnitude (for definition, see Oke 1974) corresponding 
to {\it Unit Conversion}. It is given by the following equation:
\begin{equation}
Zero~Point=(-2.5~log~(Unit~Conversion) \times (\lambda_{mean})^{2}) - 2.407
\end{equation}
where $\lambda_{mean}$ is in {\it \AA} and Unit Conversion is in $(erg~cm^{-2}~s^{-1}~\AA^{-1})$. 

\begin{figure*}[h]
\begin{center}
\includegraphics[scale=0.6]{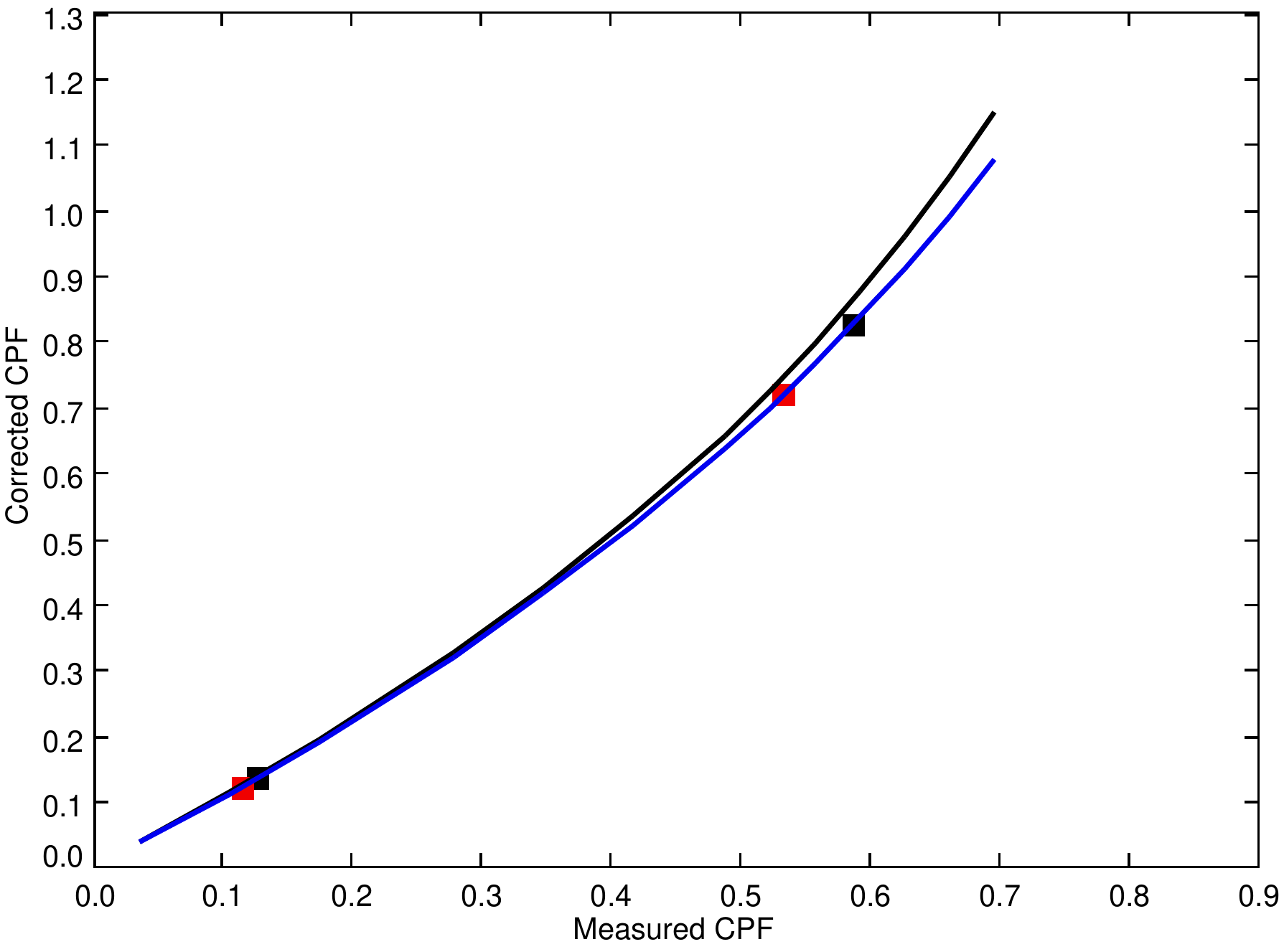}
\caption{The saturation correction for the observed Counts per frame (CPF).  
} 
\end{center}
\end{figure*}
                                                         
\begin{figure*}[h]
\begin{center}
\includegraphics[scale=0.4]{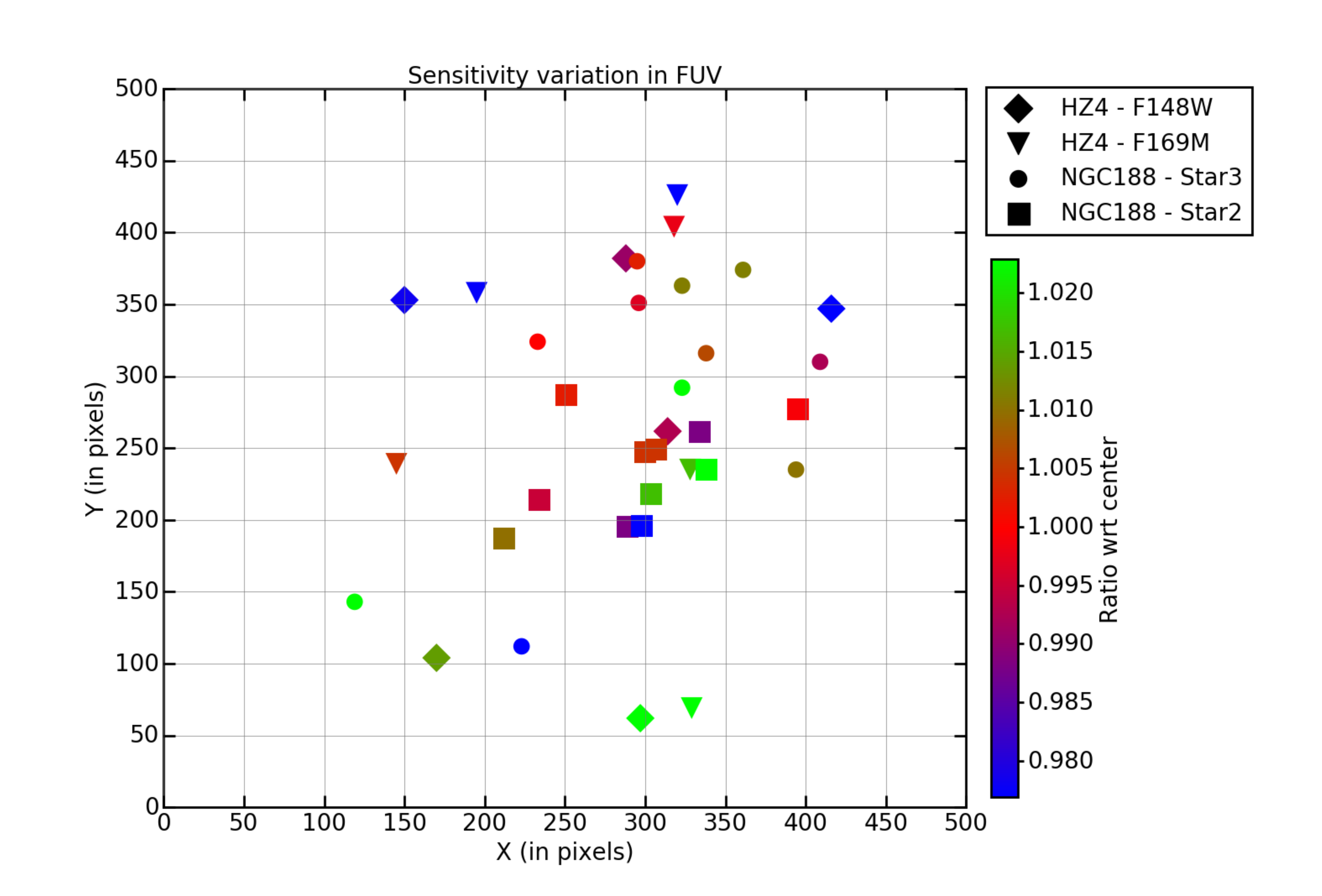}
\caption{The spatial variation of sensitivity for the FUV channel. We have shown the estimations from the standard star, HZ4, and two stars from the open cluster field, for the filters F148W and F169M. The color code is based on the ratio of the CPS at each position and the CPS of the same source at the center. } 
\end{center}
\end{figure*}

\begin{figure*}[h]
\begin{center}
\includegraphics[scale=0.7]{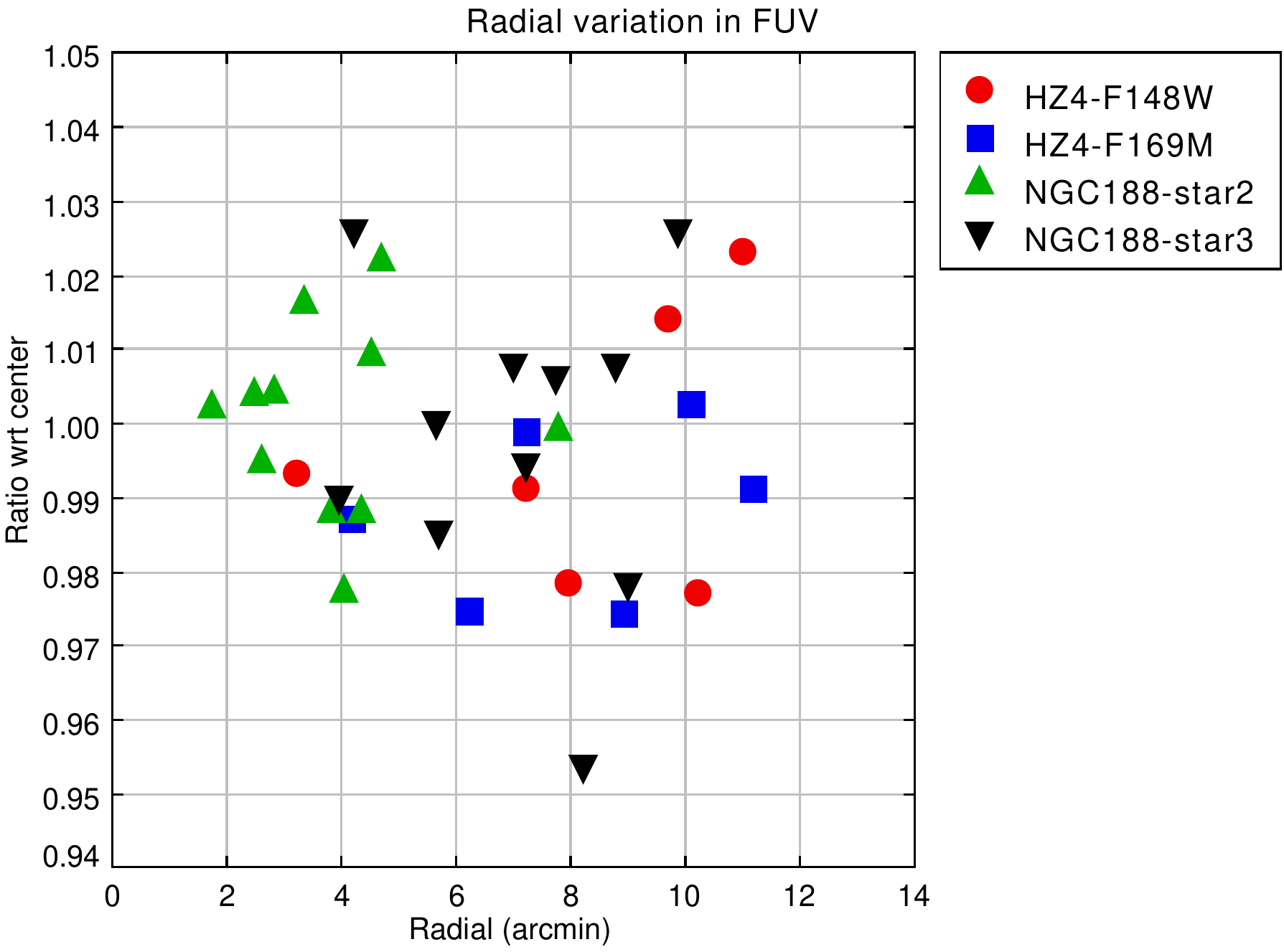}
\caption{The radial variation of sensitivity for the FUV channel.  We have shown the estimations from the standard star, HZ4, and two stars from the open cluster field, for the filters F148W and F169M. }
\end{center}
\end{figure*}

\subsubsection{Saturation and Flat Field effects}
The observations of photometric standard (HZ4) suffer from some saturation, and  in addition the photometry needs 
corrections for flat field effects. We discuss the saturation and flat field effects here.\\
{\bf Saturation:} In photon counting mode a photon event is identified by a  local peak of signal, larger than a chosen threshold, 
within a window of 3$\times$3 pixels of the CMOS imager. In case two or more photons fall on  one or  two adjacent pixels, these are counted as  a single photon. 
 Therefore, if the average photon rate for a point source is  not $<<$ 1/frame, some photons are lost in the recorded counts 
and in effect saturation results. Such saturation can be corrected for by invoking Poisson statistics for occurrence 
of photons. We have made correction for this saturation by the following equation:\\
\begin{equation}
X  =  - ln F0 
\end{equation}
where $X$ is the corrected  {\it count/frame} and $F0$  is the fraction of frames with no photon within the window 
which defines extent of the source, e.g., for  a point source the window should cover a diameter of 25\arcsec ($\sim$ 7.5 pixels) to include wings of the PSF.  
 The fraction (F0) can also be estimated from the observed counts per frame if multiple photons are not spatially separated or always occur in a window of 2$\times$2 pixels of the imager. However, the PSF extends to more than 2 pixels and some fraction of multiple photons would be spatially resolved, i.e. not fall on neighbouring pixels.  Therefore, we need some empirical method to estimate the correction for saturation from the observed counts per frame. 
In practice we have found that, for observed rates up to 0.6 per frame, the following process gives an accurate value of the correction for saturation: a) find the rate within a window of radius 7.5 pixels (25\arcsec), b) Find the correction for 97\% of the counts per frame (CPF5), as per Poisson statistics, c) The correction found is modified to get the actual correction for counts per frame. It is found that 97\% of the photons fall in the
central $5\times5$ pixel$^2$, where the saturation correction needs to be applied. This process is captured in the following equations: 
\begin{equation}
CPF5 = (1- exp(-ICPF5))\\
\end{equation}
\begin{equation}
ICORR = (ICPF5) - (CPF5)\\
\end{equation}
\begin{equation}
RCORR = ICORR \times (0.89 - 0.30 \times (ICORR)^2)\\
\end{equation}
where, ICORR is the ideal Correction for saturation, RCORR is the real correction. 
This correction is illustrated in Figure 3, for exposures with full field (28.7 frames/s). The figure shows
the measured counts per frame (CPF) on the x-axis and the corrected CPF in the y-axis. The corrected CPF using
the ideal correction in equation (12) is shown as black and the corrected CPF using the equation (13) is shown in blue. The points shown are the observations of HZ4 in two FUV filters, observed at two different frame rates, 
to get two different CPF. These observations requiring two different saturation corrections, produced the same
CPS, thereby validating the equation for saturation correction.
In addition to the saturation discussed above, for high fluxes the efficiency of detecting photons goes down as 
the signals for individual photons are reduced due to high impedance of the micro-channel-plate in the detector. 
From tests on the engineering model detector, which is similar to those used in the payload, it is inferred that for a point source 
any reduction in the efficiency of detecting photons is $<$ 5\% for 150 CPS (and $<$1\% for 30 CPS). 
This effect is ignored in the calibrations.  We note that there are no additional saturation effects relating to global count rate as each pixel of the CMOS-imager independently integrates the light falling on it.   \\
{\bf Flat field:} Any measured counts need to be corrected for variation of the sensitivity across the field or for 
flat field effect. The correction is assumed to have two distinct components of high frequency (on scales $ \lesssim$ 100 
pixels) and low frequency (on scales $ \gtrsim$ 100 pixels) respectively. The high frequency component is taken from 
the calibrations on ground done for the central wavelength and the  final low frequency component is derived from observations 
of a source at 9 points in the field (see Subsection 3.1.5). In the calibrations done on ground, transmission of all the FUV/NUV filters, 
with the exception of N219M, was found to be uniform to better than $< 0.7\%$ rms (when taken over spatial scales 
corresponding to size of the beam for a point source). Therefore, for all the filters in each of FUV and NUV, 
except for N219M in NUV, the flat field correction is expected to be nearly identical. The images are corrected for the variations in sensitivity of the detector as obtained in the ground calibrations.
The variations seen in the corrected images are taken as the low frequency variations, due to the optics-filter-detector chain, to be corrected for in the final images.

\begin{figure*}[h]
\begin{center}
\includegraphics[scale=0.35]{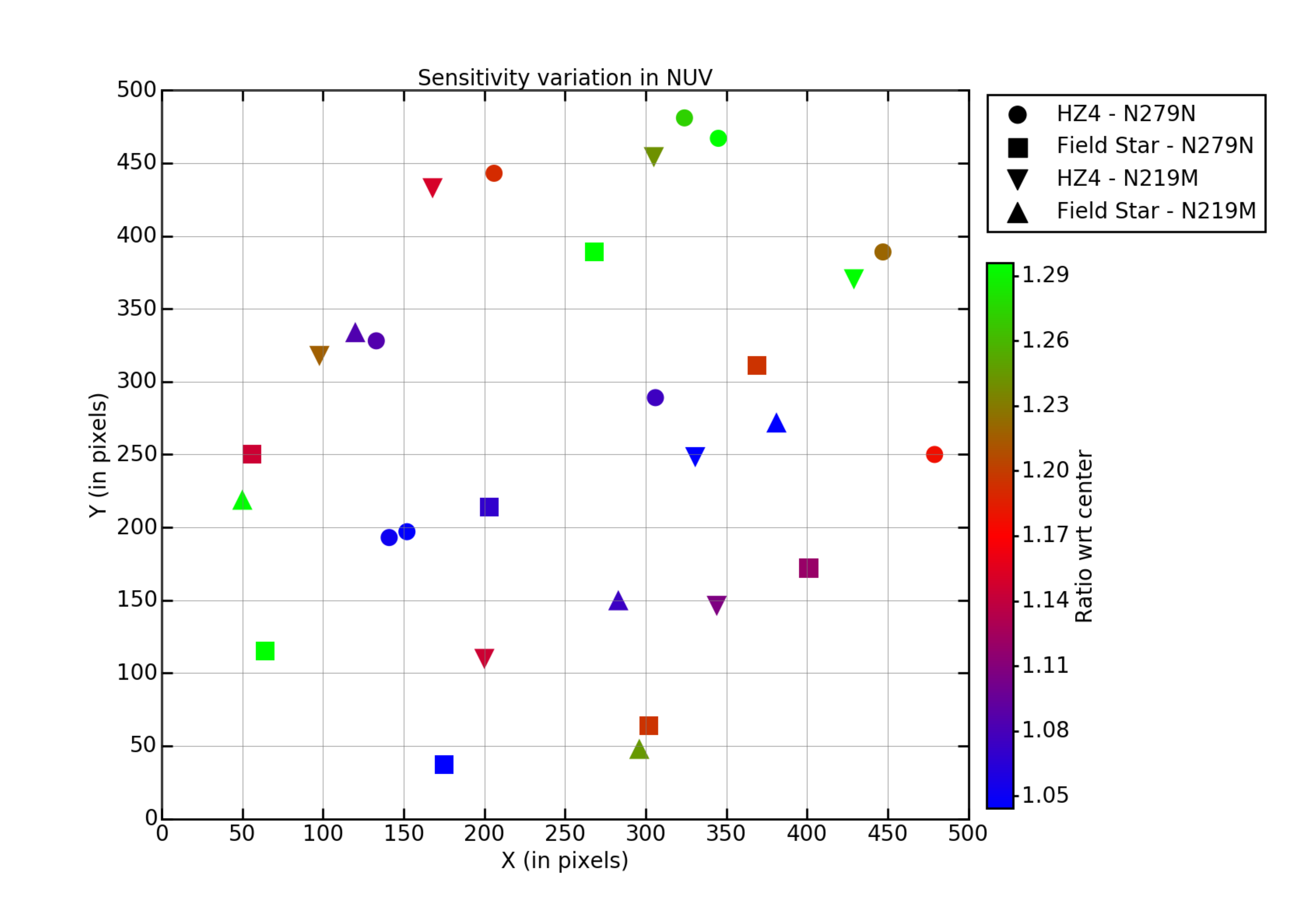}
\caption{The spatial variation of sensitivity for the NUV channel. We have shown the estimations from the standard star, HZ4, and one more bright star in the field, for the filters N219M and N279N. The color code is based on the ratio of the CPS at each position and the CPS of the same source at the center.}
\end{center}
\end{figure*}

\begin{figure*}[h]
\begin{center}
\includegraphics[scale=0.35]{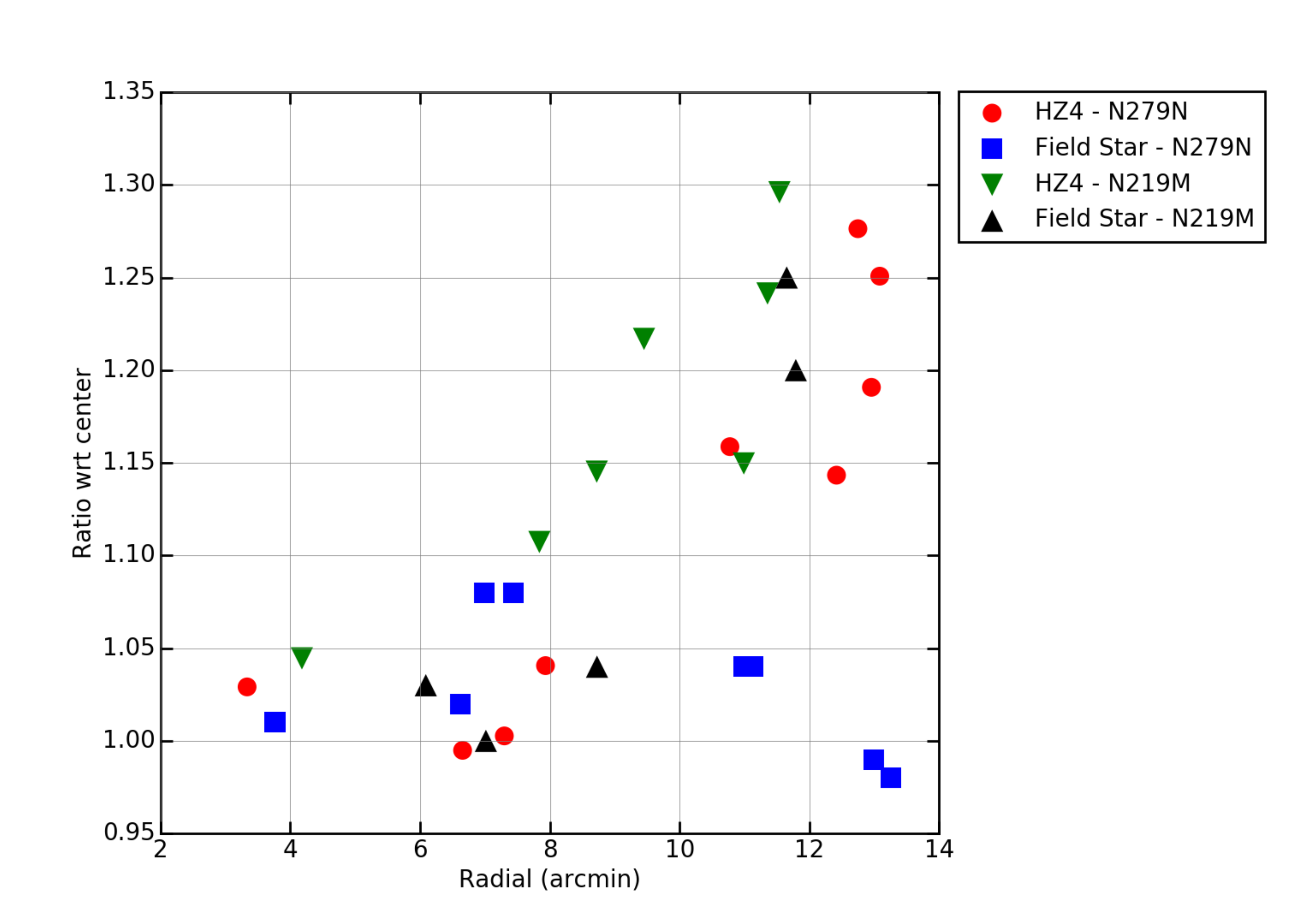}
\caption{The radial variation of sensitivity for the NUV channel. We have shown the estimations from the standard star, HZ4, and one more bright star in the field, for the filters N219M and N279N. }
\end{center}
\end{figure*}

\subsubsection{Observations of HZ4}
HZ4 gives moderate count rates in all the FUV filters, whereas it has relatively high count rates in the NUV broad band filters. We observed this
star in February 2016 and December 2016. In February 2016, which was the early period of mission operations, we used only the full
frame mode for imaging (with a read rate of $\sim$ 29 frames/sec). We observed HZ4 in all the FUV filters and only two NUV 
filters (N219M and N279N). In December 2016 , HZ4 was observed in $ 200\times200$ 
window mode (with a read rate of $\sim$ 172 frames/sec) for the F148W and F154W filters of FUV and N245M and N263M filters of NUV. Observation for the broadest NUV filter, N242Wa, was done with the $150\times$150 window (with a
read rate of $\sim$ 292 frames/sec). All these data were used to estimate the final calibration values.Typical S/N for these data is 50 to 100, and the inferred rates for exposures with full frame and with $ 200\times200$ window agree within the errors which provides empirical verification for the process of correcting saturation. 
 We note that the readout with rolling-shutter action splits a small fraction (1/number of rows in the frame) of events between two consecutive frames. 
Each of such events on average contributes 1.5 counts. Any errors due to this effect is always $<$ 0.5\% and no correction is made for it.

HZ4 was also observed in 9 different positions on the detector to estimate the variation of sensitivity across the field.
The observations were made in the center, and then followed by 8 points separated by 45$^o$, along a circle of radius 10\arcmin.
The observations were done in February 2016 using the filter F169M in the FUV channel and N279N in the NUV channel. The observations
were repeated in December 2016, for the F148W filter in the FUV and the N219M filter in the NUV. Even though all the 9 locations in the detector
were observed,  due to reasons such as, space craft drift, problems with the data etc, we could not retrieve all the observations. We used only those observations which had SNR more than 30. 

\subsubsection{Estimation of Zero Points and Unit Conversions}
Images were generated, after correcting for flat-field (as obtained in the ground calibrations), distortion (see Subsection 3.4) and drift of pointing, by the stand alone software CCDLAB (Postma et al. 2017, in preparation). Images were also generated without correcting for flat-field. Aperture photometry was performed on both the sets of images using IRAF and DAOPHOT and we estimated the total flux within a radius of 7.5 pixels (about 25\arcsec). The background contribution was subtracted from an outer annulus. The estimated CPS (counts per second) was corrected for saturation as described in subsection 3.1.2. The correction for saturation is performed on the counts in those images which are not corrected for flat-field. The saturation corrected value of the counts is then corrected for the 
sensitivity variation, which is provided by the image corrected for flat-field. This correction factor is estimated as the ratio of the observed counts in the flat corrected image to the counts from the un-corrected image. The final corrected counts are used to derive the Unit Conversion factors and the Zero Point magnitudes. The results are shown in Table~4.  
 The initial estimations of the Unit Conversion factors were used by Subramaniam et al. (2016b), for estimating the flux. The results presented here are the revised estimations, when compared to their Table 1. Their study demonstrated that the flux estimated from UVIT are found to be in good agreement with those estimated from GALEX, UVOT and UIT within errors. 

\begin{table*}[h]
\begin{center}
\caption{Performance parameters with individual filters are shown for the FUV and NUV Channels. Here ZP and UC refer to Zero Point magnitude and Unit Conversion respectively. }
\begin{tabular}{lccccccccc}
\small
Filter Name & CEA$^a$& EEA& Err& ZP$^a$ & ZP&Err& UC$^a$ & UC & Err \\
\hline
FUV:&&&&&&&&&\\
\hline
F148W&10.50&8.70&0.08&18.221&18.016&0.01&2.56E-15&3.09E-15&2.9E-17\\
F148Wa&9.94&8.16&0.06&18.158&17.994&0.01&2.69E-15&3.28E-15&2.5E-17\\
F154W&11.46&9.55&0.11&17.975&17.778&0.01&2.961E-15&3.55E-15&4.0E-17\\
F172M&8.96&8.62&0.13&16.383&16.342&0.02&1.03E-14&1.074E-14&1.6E-16\\
F169M&10.27&9.70&0.08&17.517&17.455&0.01&4.15E-15&4.392E-15&3.7E-17\\
\hline
NUV:&&&&&&&&&\\
\hline
N242W&56.01&47.21&0.13&19.996&19.81&0.002&1.87E-16&2.220E-16&6.5E-19\\
N245M&48.84&40.01&0.19&18.715&18.50&0.07&5.94E-16&7.25E-16&3.6E-18\\
N263M& 37.84&32.52&0.36&18.339&18.18&0.01&7.26E-16&8.44E-16&9.6E-18\\
N219M&12.31&6.39&0.10&17.297&16.59&0.02&2.72E-15&5.25E-15&8.2E-17\\
N279N& 24.55&22.58&0.22&16.593&16.50&0.01&3.22E-15&3.50E-15&3.5E-17\\
\hline
\end{tabular}
\end{center}
\vskip 1.0ex
\small{
{\bf Notes:} $^{a}$ Measurement from ground calibrations.  The differences between the values of CEA and EEA are primarily due to uncertainties in the ground calibrations, and values of EEA are to be used for analysing the images.\\
}
\end{table*}

\subsubsection{Low Frequency Flat Field Corrections}
The observations on HZ4 at various positions on the detector was used to estimate the variation of sensitivity across the 
detector. The observations of the 9-points were used to estimate the 
CPS of HZ4 on various locations in the detector, before and after flat field correction. In order to get a better coverage, we also used two bright FUV stars in the field of NGC 188. This cluster was observed 
every month during the performance verification phase and once every three months after that. The FUV bright stars in this field are used to track the sensitivity of the FUV channel. Including these 11 observations of two stars helped to increase the coverage of the observed field to understand the sensitivity variation in the FUV channel.
Any variations in the
CPS (estimated after flat correction), across the detector, would indicate that the flat field as estimated in the ground calibrations
needs modification. The detected variation could be modelled as a low frequency variation and convolved with the ground flat field. We first discuss the variation in the FUV channel.\\
{\bf FUV Channel:} We have observed HZ4 in F148W and F169M filters across the field. In addition NGC188 was observed in F148W filter. As these filters do not have any significant 
variation in transmission across the filter, we assume that all the detected variations are due to the detector. The values of 
the corrected CPS have been normalised with CPS near the centre (used to derive the parameters in Table 3). The variation 
of normalised values is shown in Figures 4 and 5. These figures suggest
that the variation in the corrected counts is within about 3 percent, or no more than twice the rms error on the counts, within a radius of 11\arcmin. There are no data on point sources to give information on
the variations for larger radii. However, by comparing the background for the same part of sky observed at different locations on the detector it is seen that counts could be overestimated by up to 15 percent near the edges.

{\bf NUV Channel:} The variation across the NUV channel is estimated using the N279N and N219M filters. We used the 
measurements of HZ4 as well as a bright field star to get better coverage.
In the ground calibrations the N279N filter did not show
any significant variation in transmission across its face, but the N219M filter showed a lot of variation and its transmission reduced by a factor two in the orbit. The values of 
the corrected CPS have been normalised with CPS at the centre (used to derive the parameters in Table 4). The variation 
of normalised values is shown in Figures 6 and 7. 
 We note that the large variations seen are unlikely to be the effects of saturation as the corrected count rates for these filters are $\sim$ 7 CPS,  which is much less than the corrected rate 
of $\sim$ 23 CPS for the filter F148W which shows very little variations over the field.  These figures suggest
that the variation in the corrected counts is less than 15 percent within a radius of 11\arcmin. However, outside the radius of 11\arcmin, the variations are seen up to about 30 percent. The largest variation is seen for locations near the top-right region shown in
figure 6, whereas diametrically opposite locations at the bottom-left side, do not show large variation. This contrast is reflected
in figure 7, where the ratio at large radii show small as well as large deviation. We plan to do more observations to accurately
estimate the sensitivity variation at large radii.

\subsection{Spectral Calibrations}

\begin{figure*}[h]
\begin{center}\includegraphics[scale=0.8]{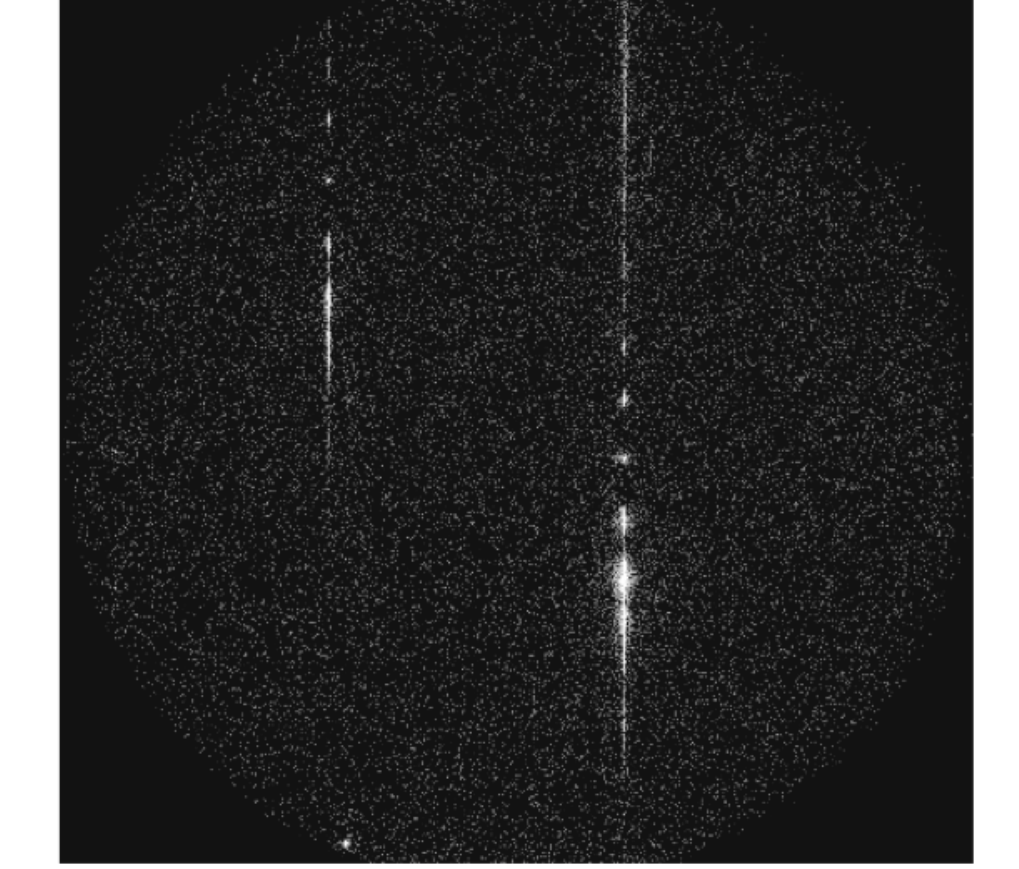}
\caption{Image of the spectrum for NGC40  obtained using the first FUV grating (\# 63771).
In the image, the brightest trail is for NGC40. The sharp point near the centre of trail is zero order from where the dispersion is counted and the second (blazed) order is below the zero order. 
}
\end{center}
\end{figure*}
\begin{figure*}[h]
\begin{center}
\includegraphics[scale=0.8]{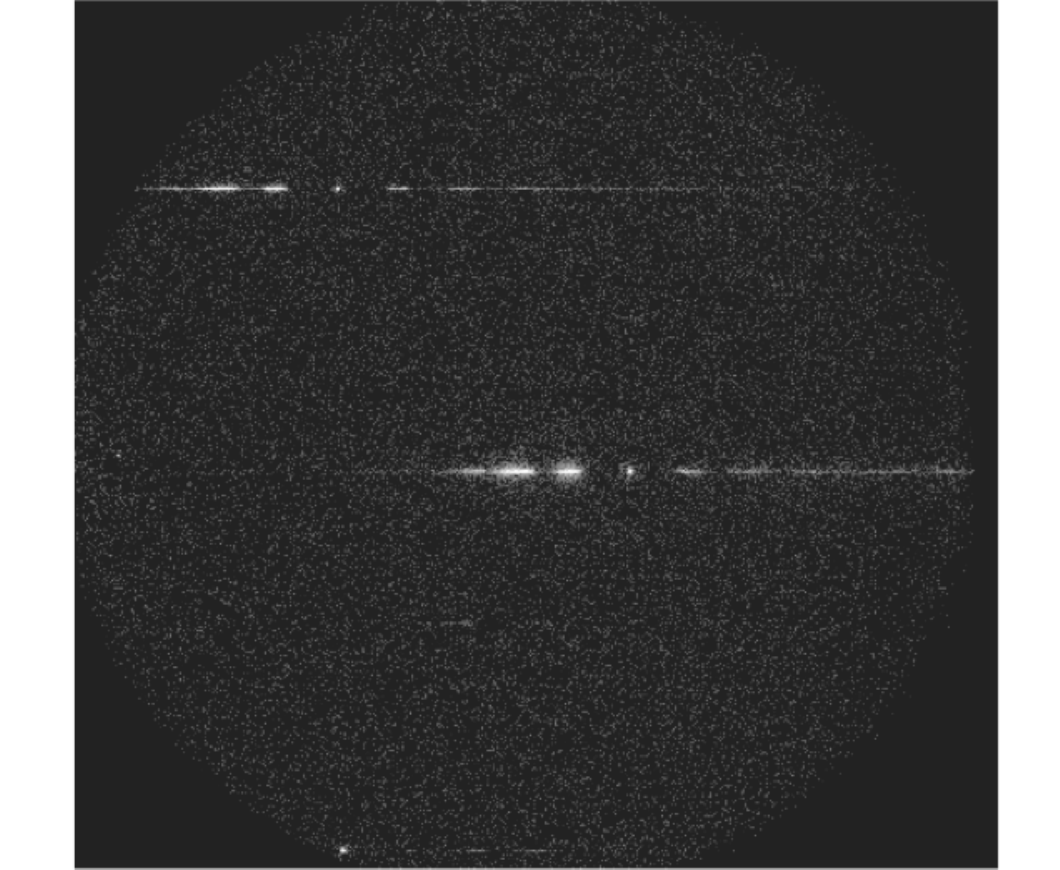}
\caption{Images of the spectrum for NGC40  obtained using the second FUV grating (\# 66126).The sharp point near the 
centre is the zero order and the second (blazed) order is on left of the zero order. 
}
\end{center}
\end{figure*}
\begin{figure*}[h]
\begin{center}
\includegraphics[scale=0.8]{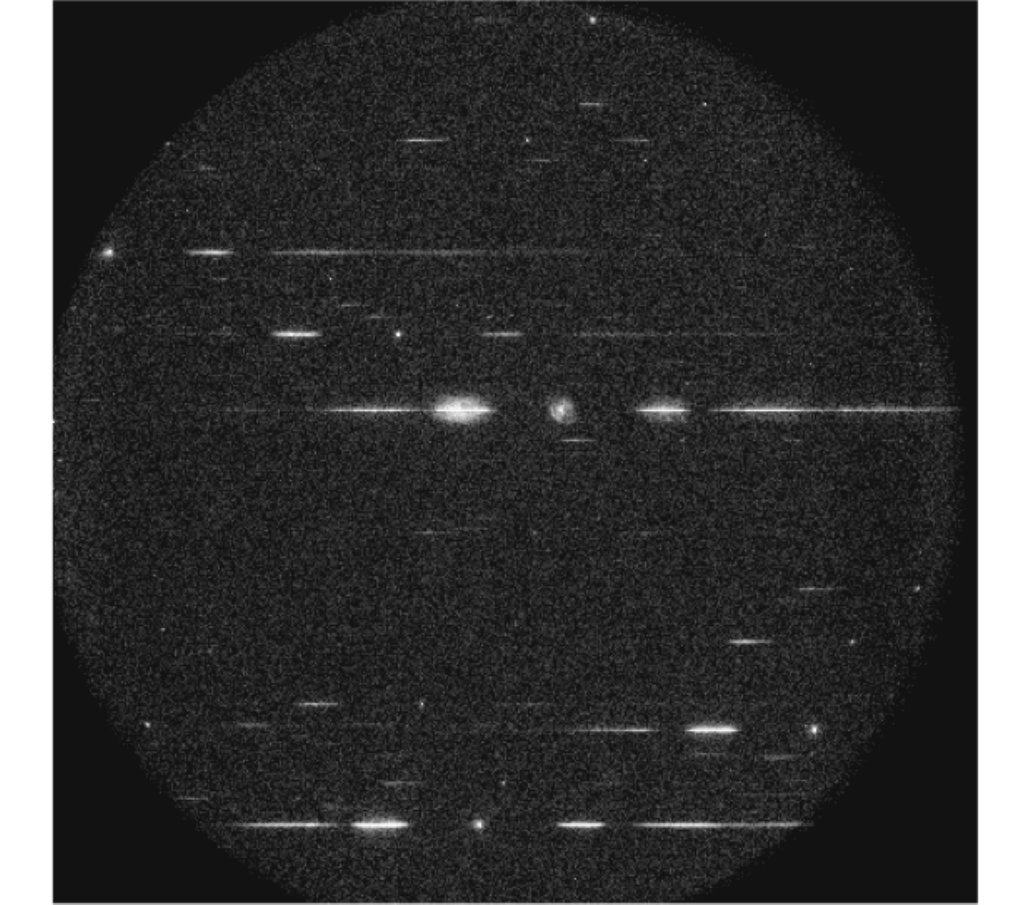}
\caption{Images of the spectrum for NGC40  obtained using the NUV grating (\# 66125). The sharp point near the 
centre is the zero order and the first (blazed) order is on left of the zero order.  
}
\end{center}
\end{figure*}

Slitless spectroscopy is implemented with gratings ruled with 400 lines/mm on CaF2 substrate. Two gratings are mounted in the filter wheel for FUV such that their dispersions are relatively orthogonal, and one grating is mounted in the filter wheel for NUV.  Planetary nebula NGC40 is used for calibrating dispersion and HZ4 is used for calibrating effective area as a function of wavelength. These calibrations are described below. For more details of these calibrations the reader is referred to UVIT In-Orbit Spectroscopy Wavelength /Flux calibration (Sriram  et al. 2017 under preparation).

\subsubsection{Dispersion and Resolution}
Dispersion is estimated by comparing the observed spectrum of NGC40 with its spectrum from IUE. The process of calibration is explained below. Images of the observed spectra are shown in Figures 8, 9 and 10. The dispersion relations are obtained by assigning wavelengths to the prominent bright features in these spectra by comparing with the respective IUE-spectrum from Feibelman (1999). The dispersion relations obtained are:  
\begin{itemize}
\item For the second order in FUV grating (\# 63771) for 1300 \AA  ~to 1800 \AA :
\begin{equation}
 \lambda(A) = -5.544 X + 53.4 (\pm 3),
\end{equation}
\begin{equation}
Spectral resolution = FWHM/dispersion \sim 5.544\times3 \sim  17 \AA;
\end{equation}
\item For the second order in FUV grating (\# 66126):
\begin{equation}
 \lambda (A) =  -5.719 X + 5.0 (\pm 3),
\end{equation}
\begin{equation}
Spectral resolution = FWHM/dispersion \sim 5.719\times3 \sim 17 \AA;
\end{equation}

\item For the first order in NUV grating (\# 66125): 
\begin{equation}
\lambda (A) =  -11.04 X + 48  (\pm 6)
\end{equation}
\begin{equation}
Spectral resolution = FWHM/dispersion  \sim 11.04\times3 \sim  33 \AA;
\end{equation}
\end{itemize}

where {\it X} is separation from the zero order in pixels.


\begin{figure*}[h]
\begin{center}
\includegraphics[scale=0.7]{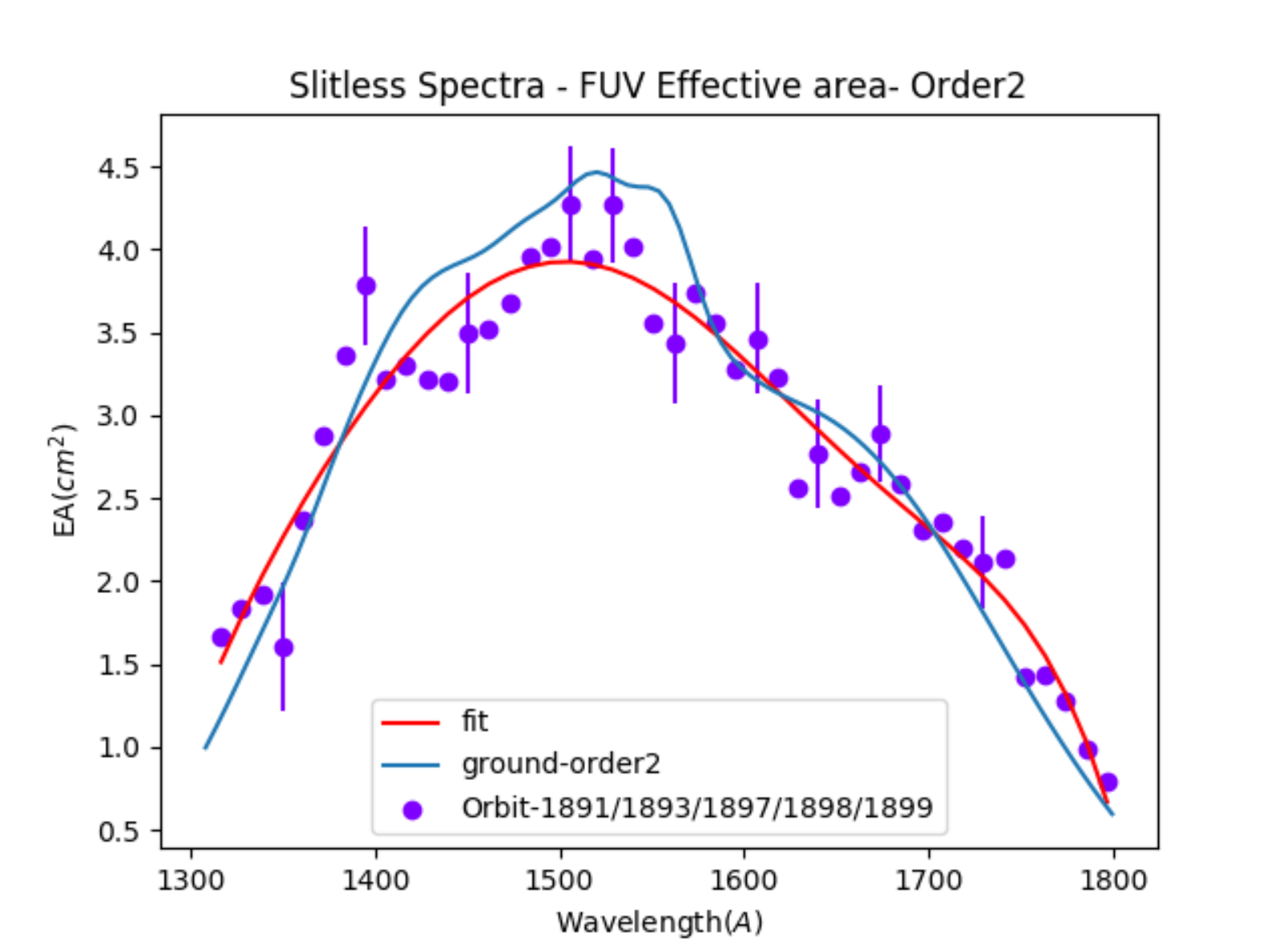}
\caption{Effective areas as a function of wavelength for the first FUV Grating.
}
\end{center}
\end{figure*}

\begin{figure*}[h]
\begin{center}
\includegraphics[scale=0.7]{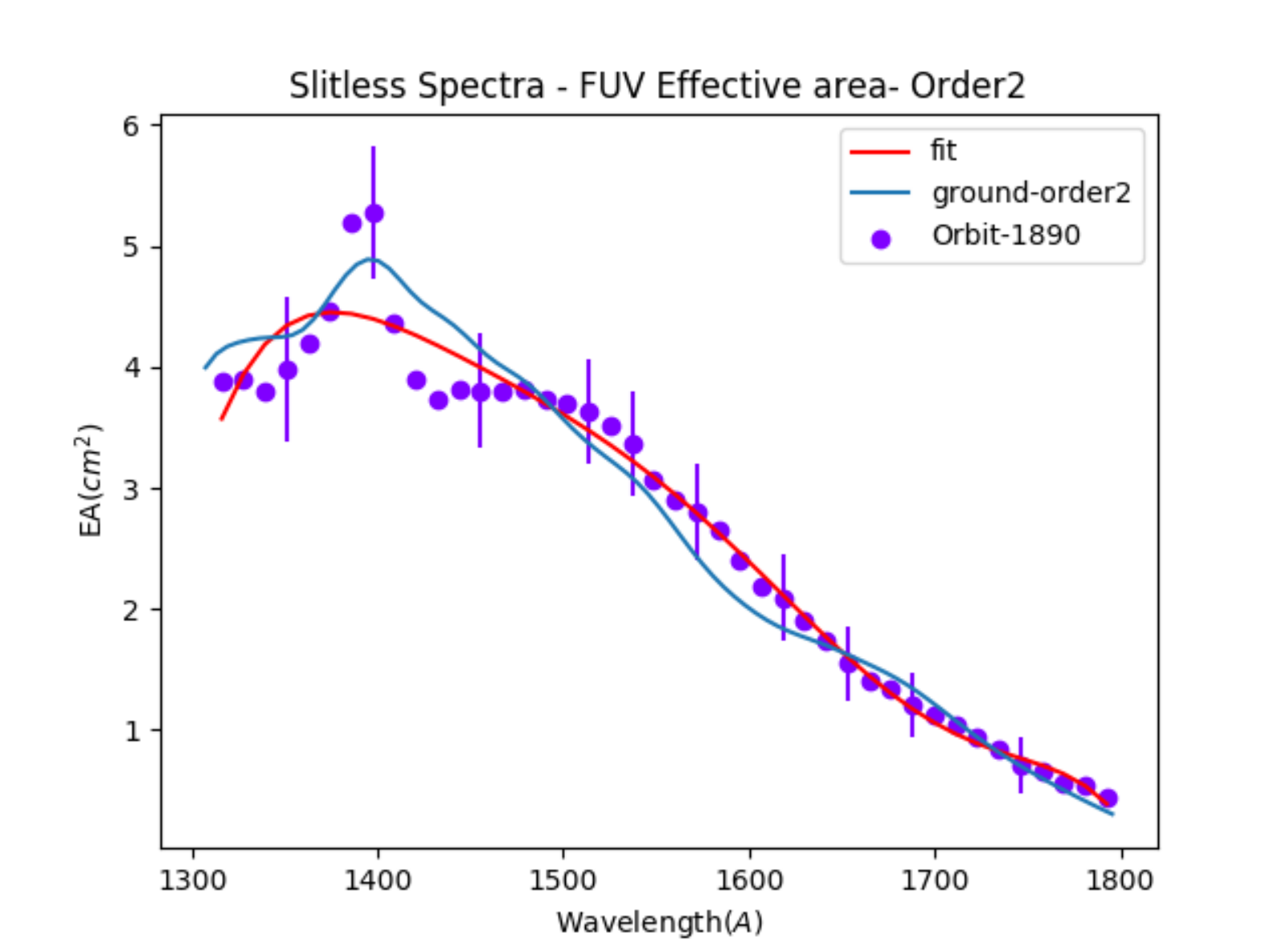}
\caption{Effective areas as a function of wavelength for the second FUV Grating.
}
\end{center}
\end{figure*}

\begin{figure*}[h]
\begin{center}
\includegraphics[scale=0.7]{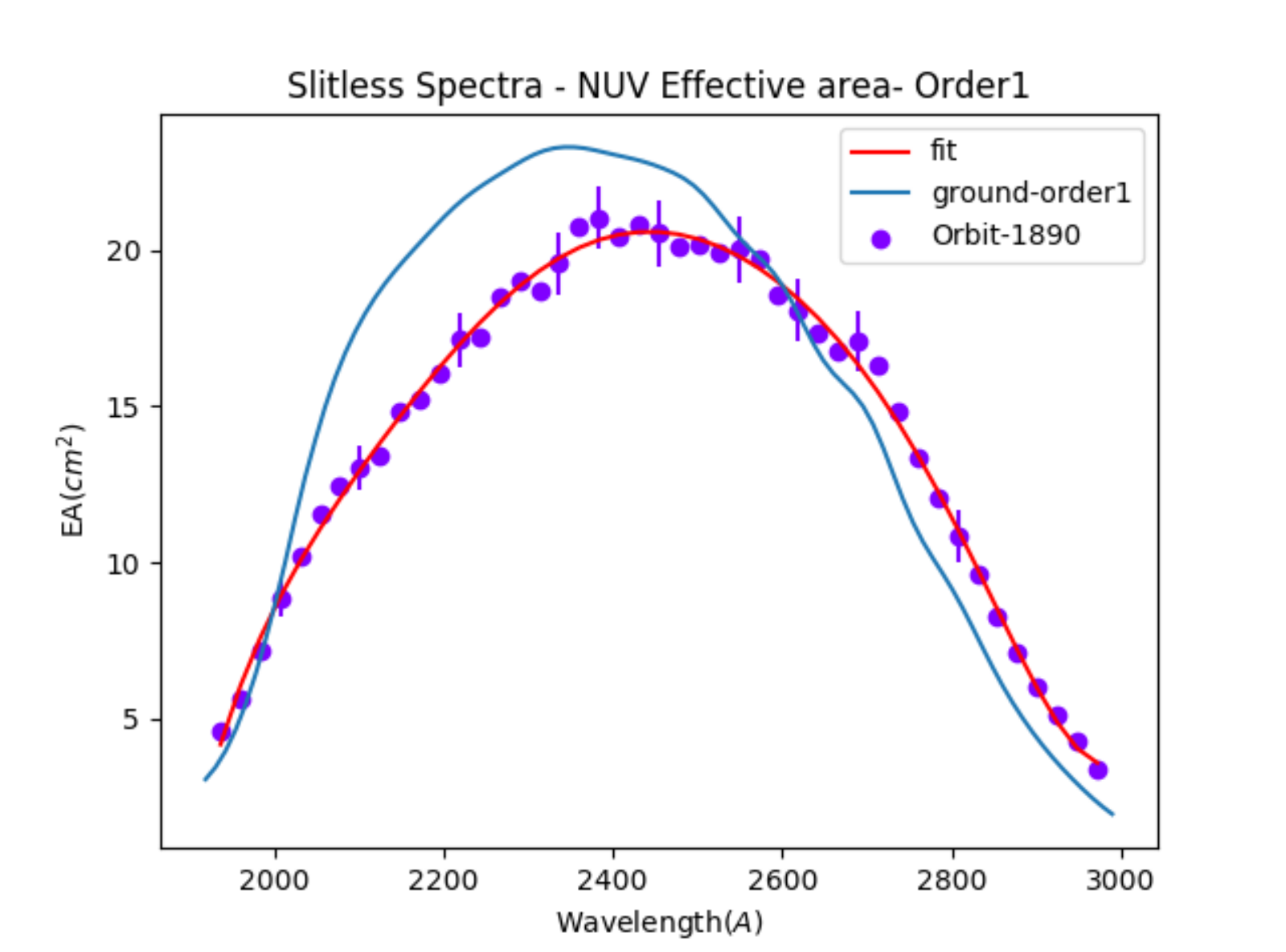}
\caption{Effective areas as a function of wavelength for the NUV Grating.
}
\end{center}
\end{figure*}
\subsubsection{Effective Area}
The calibrations for effective area, shown in Figures 11, 12 and 13, are done with observations of the photometric standard HZ4. The estimates based on the ground calibrations for transmissions of the gratings are shown as red lines, and errors on the results of observations are only shown for a few wavelengths to avoid crowding. The results are in satisfactory agreement with the calibrations done on ground. We note that a common cause for lower values of the effective area (at 2000-2400 \AA) from in-orbit  calibration of the grating and N219M filter  for NUV (see Table 3) could be due to an overestimate of either the quantum efficiency of the detector or transmission/reflectivity of the optical elements in the ground calibrations.  

\subsection{Point Spread Function}
The point spread function (PSF) for UVIT not only depends on quality of the telescope-optics and the intrinsic spatial resolution of the detector, it also depends on some secondary factors. The secondary factors are: i) lack of adequate thermal control of the telescope which could
lead to defocus, ii) distortions in the detector, iii) Inclination of the detector plane, and curvature of the focal plane, and iv) any leftover errors in the estimated drift used in shift and add algorithm to combine the large number of short exposures. In addition to these factors, for the filter NUV-B15 there are additional errors due to optical imperfections in this filter (which is made by gluing three pieces).  It is found that the thermal control is very stable and is not expected to lead to any temporal variations in the focus. Therefore, temporal variations are not expected in the PSF. 
Images of a part of SMC are selected to study the PSF as this  region has stars densely covering the field of view. We analyse the PSF in two parts: i) the central core, ii) the extended wings. The extended wings are caused primarily by scattering on surface of the mirrors and due to the supporting ribs which block the aperture. Any variability over the field is only expected for the central core, while the wings are not expected to show such variability. Therefore, we only present the variability for the central core. In those cases where the corrections for drift of the S/C are not optimal, the central core would broaden.

\begin{figure*}[h]
\begin{center}
\includegraphics[scale=0.7]{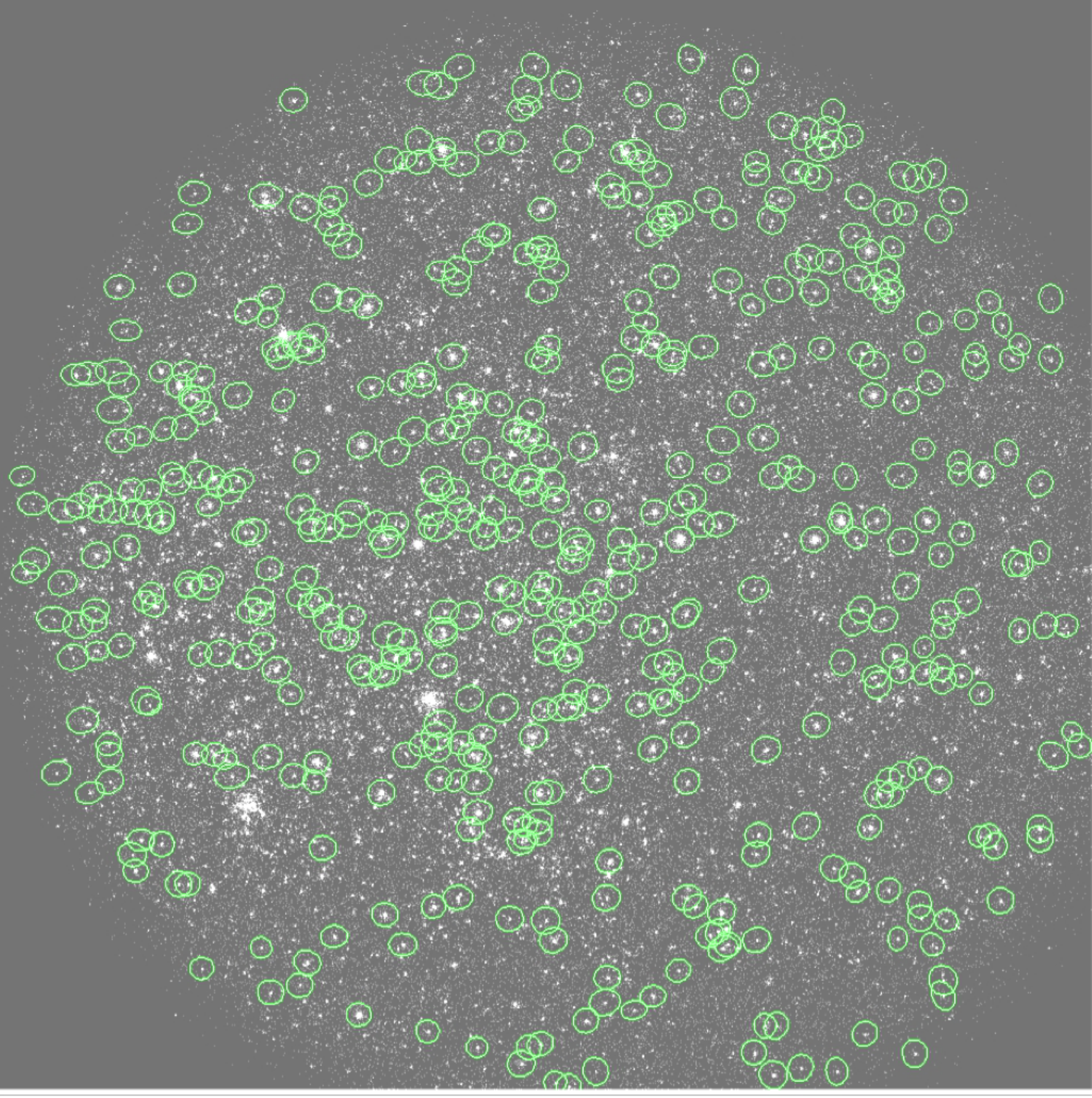}
\caption{Variation of PSF over the field for FUV (F148W) is shown with green ellipses representing FWHM along 
two directions. For FUV, the FWHM varies  little over the field. 
}
\end{center}
\end{figure*}

\begin{figure*}[h]
\begin{center}
\includegraphics[scale=0.7]{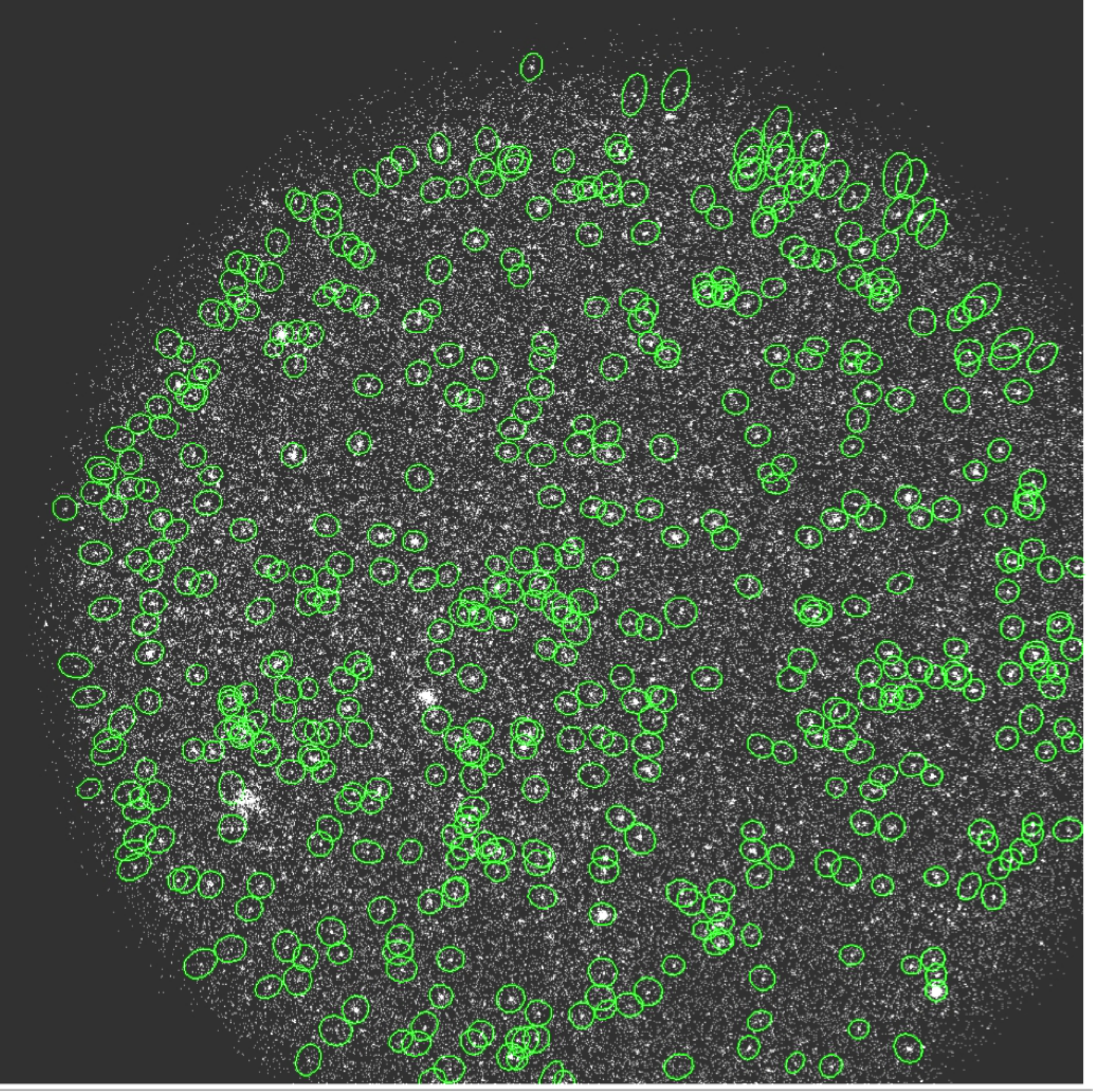}
\caption{Variation of PSF over the field for NUV (N279N) is shown with green ellipses representing FWHM 
along two directions. For NUV, FWHM and ellipticity are large near top-right corner: this seems to be due 
to remainders of the distortion.
}
\end{center}
\end{figure*}

The Central Core: The inner 9\arcsec ~diameter of the PSF is fitted to circular Moffat function. Average FWHM is found as $\sim$  1.26 $\pm$ 0.15\arcsec ~for the NUV image with filter N279N and as $\sim$  1.31 $\pm$ 0.10\arcsec for the FUV image with filter F148W. The FWHM is larger near the edges, and can be as large as 2.1\arcsec ~at some parts. Ellipticity is typically less than 0.1 but can he as high as 0.3 near the edges. As the optical path is matched for all the filters of each channel, the FWHM is expected to be similar for all filters, with the exception of filter N219M of NUV which has a poor optical quality and gives an average FWHM of $\sim$ 1.9\arcsec. Variation of PSF over the field is shown in Figures 14 and 15.  The variations for  FUV are small, while NUV shows an increase of $\sim$ 10\% in FWHM in the central part of the field compared to the edges, which could be due to non-optimal placing of the detector (photo-cathode) combined with curvature of the focal plane. The NUV images also show large ellipticity (near the edges) at the same locations where astrometric errors are large, and remainders in distortion-correction seem to be cause of both.  The percentage of the counts as a function of radial distance is tabulated in Table~5. The radial growth curves are shown in Figures 16 and 17.  About 10\% energy is lost in the pedastal.  We also caution that for the central parts of the PSF, say for radius $<$ 2\arcsec, the shape would depend on the perturbations in tracking the aspect and small variations of the focus. 

\begin{figure*}[h]
\begin{center}
\includegraphics[scale=0.7]{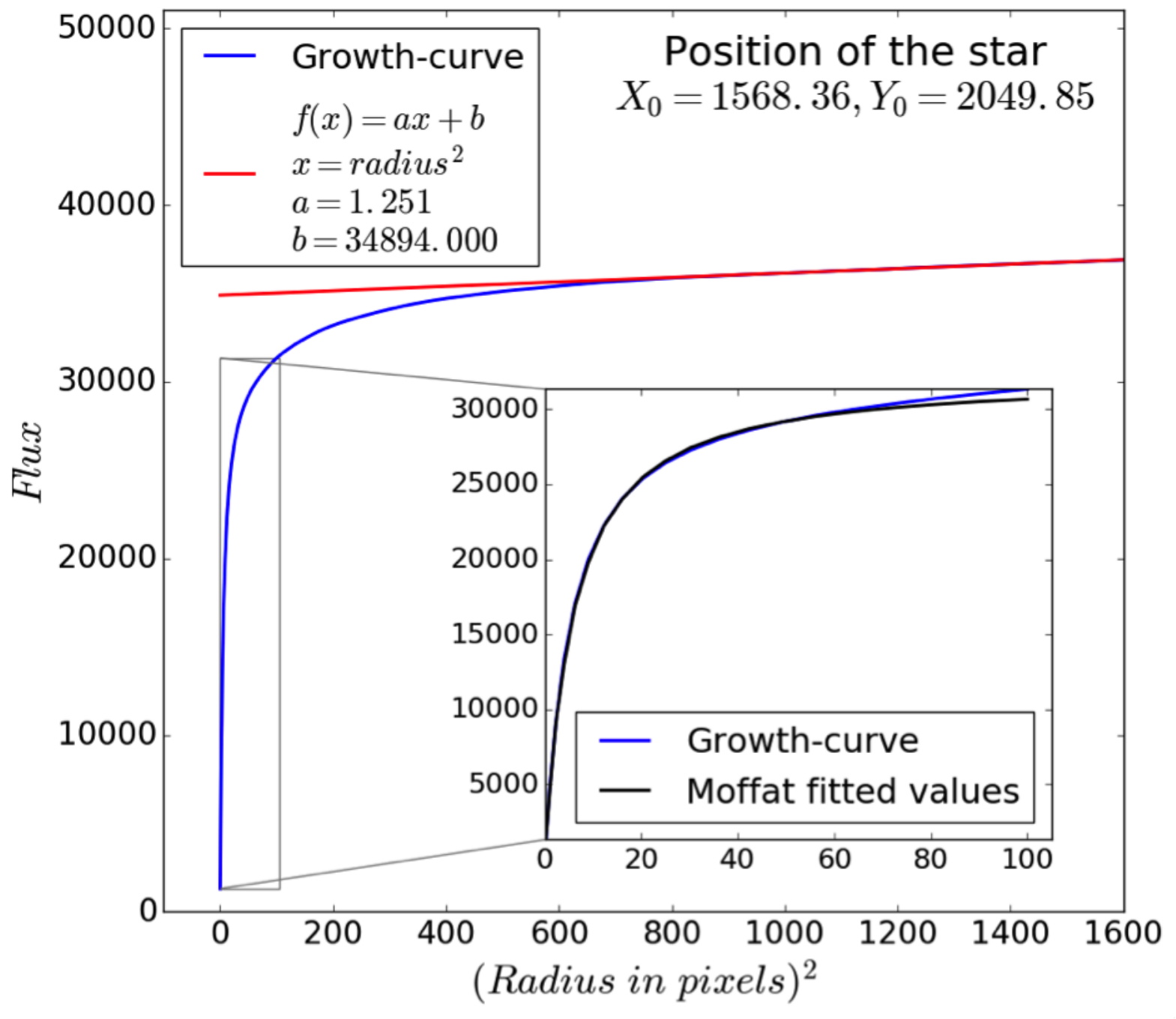}
\caption{Growth curves for the PSF FUV (F148W). The flux is in units of total photon-counts. The inset at top-left shows the equation used to fit the background by fitting the points between radii of 30 and 40 pixels. 
}
\end{center}
\end{figure*}

\begin{figure*}[h]
\begin{center}
\includegraphics[scale=0.7]{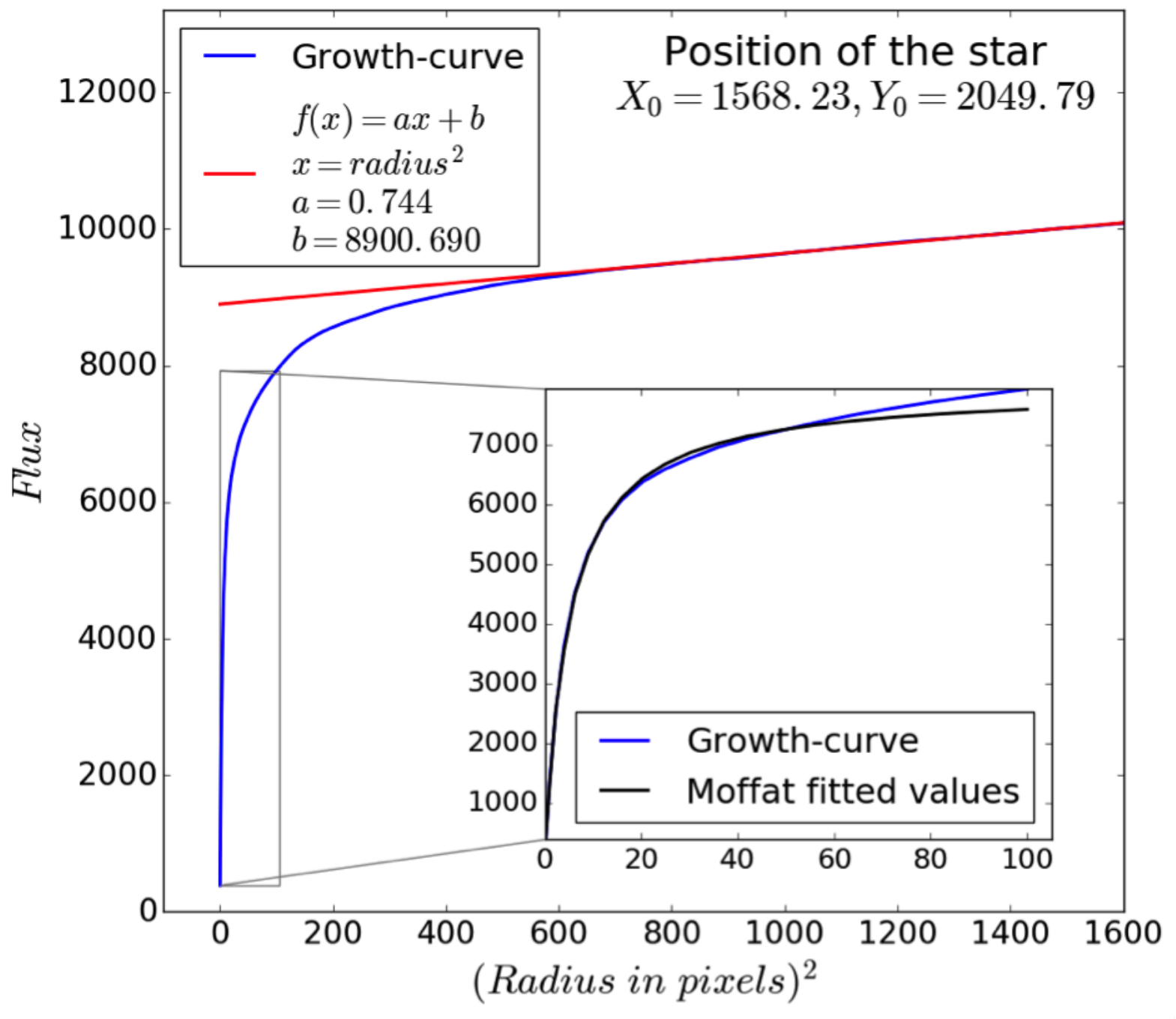}
\caption{Growth curves for the PSF for NUV(N279N).  The flux is in units of total photon-counts. The inset at top-left shows the equation used to fit the background by fitting the points between radii of 30 and 40 pixels. 
}
\end{center}
\end{figure*}
\subsection{Astrometry}
The detectors of UVIT have distortion, i.e. the recorded positions on the CMOS imager are not transformed to the focal plane through a linear relation. The distortions were calibrated on ground by imaging a grid of regularly spaced holes, and the results are used to translate the positions from the CMOS imager to the focal plane (see Girish et al 2017) . The final astrometric accuracy obtained is checked by comparing stellar positions in the FUV and NUV images of SMC obtained with UVIT, as well as by comparing the positions in UVIT images with the positions DSS-blue image.  In each comparison the relative gain is floated to get the best fit. The results of these comparisons are presented in Figures 18, 19, 20 and 21, and represent upper limits on the errors in UVIT images.  As FUV and NUV do not share any optical element, and the telescopes (of identical design) give $<$ 0.2\arcsec ~distortion which is modelled, we consider that the inter-comparison of the two channels shown in Figure 17 represent the best estimate of the astrometric errors for UVIT, i.e. an rms of less than 0.5\arcsec.

\begin{figure*}[h]
\begin{center}
\includegraphics[scale=1.0]{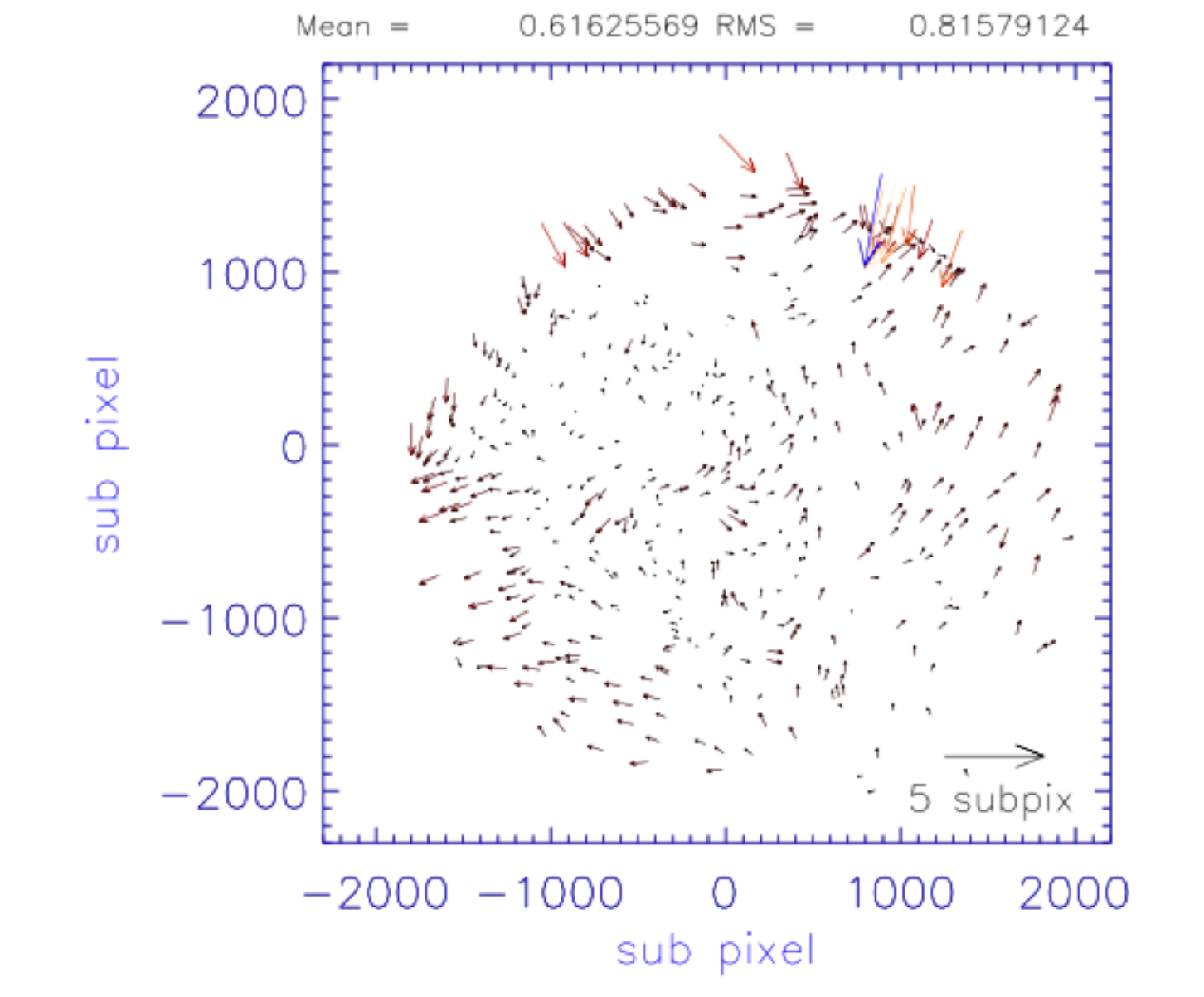}
\caption{Relative differences in positions  of the stars in the UVIT images taken with FUV (F148W) and NUV (N279N) are shown as vectors, where tail of the vector corresponds 
to position in the field. Positions and errors are shown in units of sub-pixel which is equal to $\sim$ 0.41\arcsec. 
We consider mean error (0.6 pixels or 0.25\arcsec) as the best estimate of relative astrometric errors in the images of FUV and NUV with all the filters except those with N219 in NUV.
}
\end{center}
\end{figure*}

\begin{figure*}[h]
\begin{center}
\includegraphics[scale=1.0]{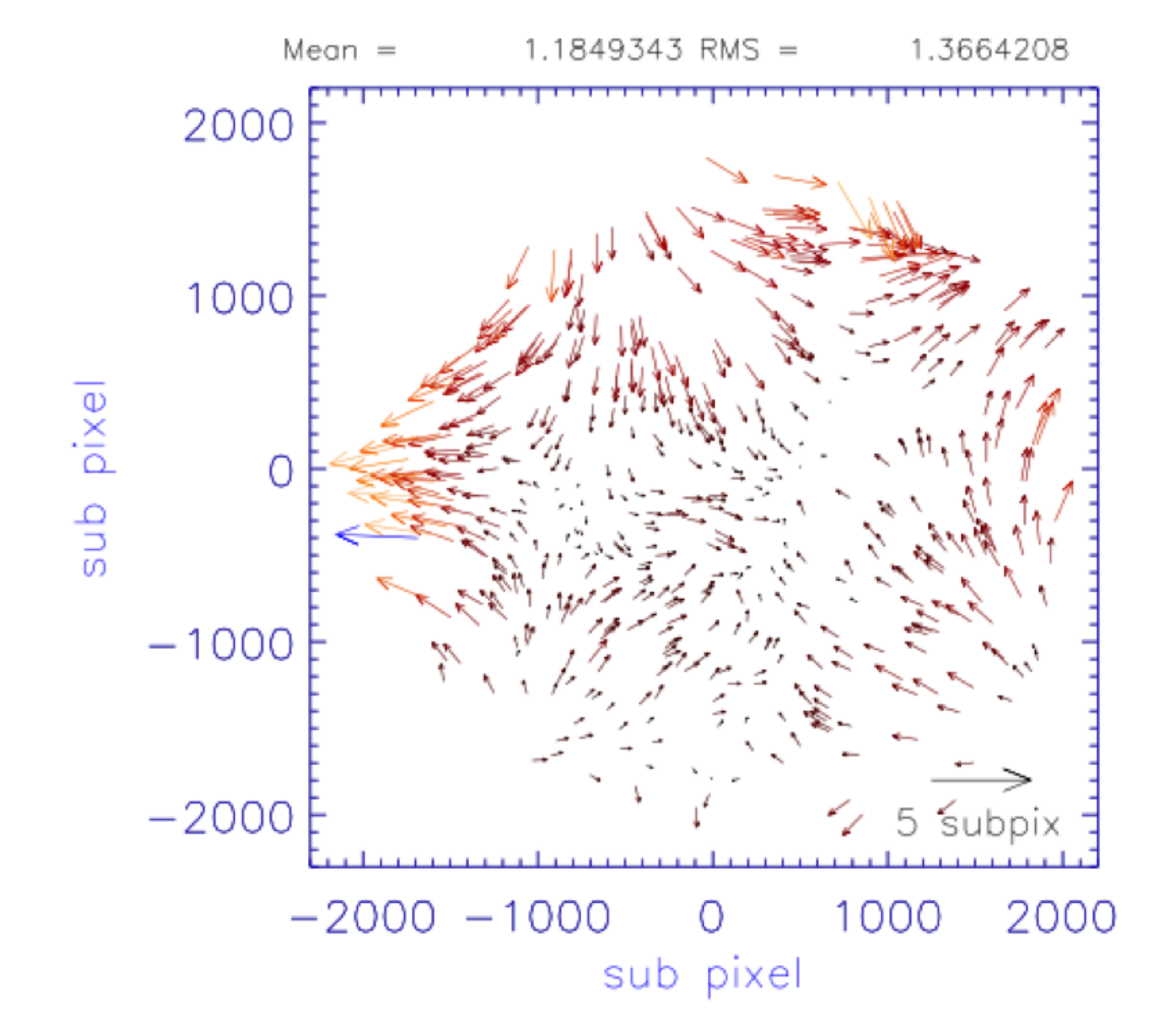}
\caption{Relative differences in positions  of the stars in the UVIT images taken with FUV (F148W) and NUV (N219M)  are shown as vectors, where tail of the vector corresponds to position in the field. Positions and errors are shown in units of sub-pixel which is equal to $\sim$ 0.41\arcsec.  
}
\end{center}
\end{figure*}

\begin{figure*}[h]
\begin{center}
\includegraphics[scale=1.0]{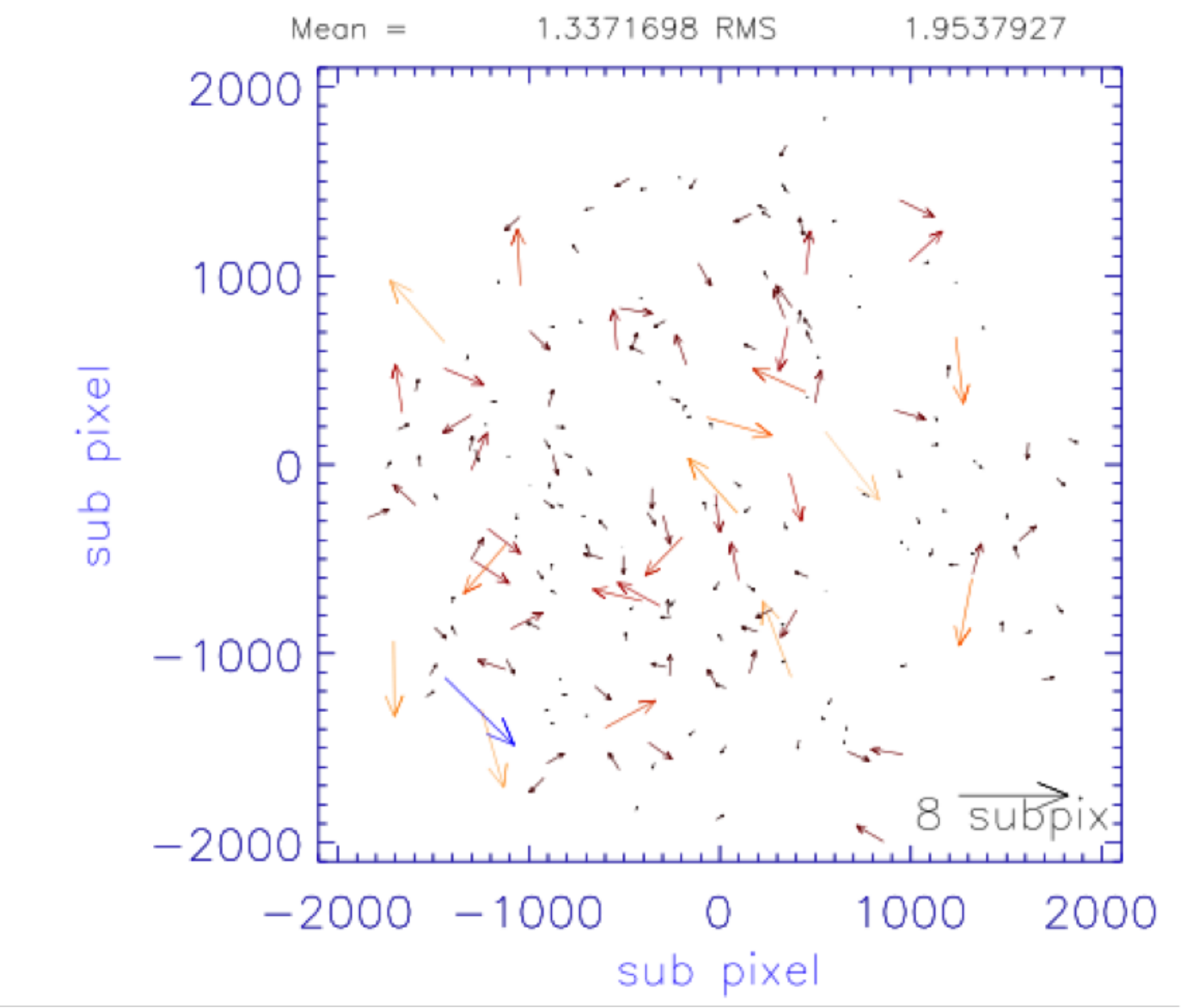}
\caption{Relative differences in positions  of the stars in the UVIT image taken with FUV (F148W) and the DSS-blue image are shown as vectors, where tail of the vector corresponds to position in the field. Positions and errors  (shown on the top)   are shown in units of sub-pixel which is equal to $\sim$ 0.41\arcsec.  
 More than 40\% of stars in UVIT images could be matched with the selected bright stars ( a total of 1025 over 45\arcmin$\times$45\arcmin) of DSS within a distance of 8 sub-pixels (3.3\arcsec). The average deviation is 1.3 sub-pixels (0.5\arcsec), but some of the stars show large deviations. On checking with the DSS image, it is found that each and every match with deviation $>$ 4 sub-pixels (1.7\arcsec) corresponds to overlapping stars in DSS. 
}
\end{center}
\end{figure*}

\begin{figure*}[h]
\begin{center}
\includegraphics[scale=1.0]{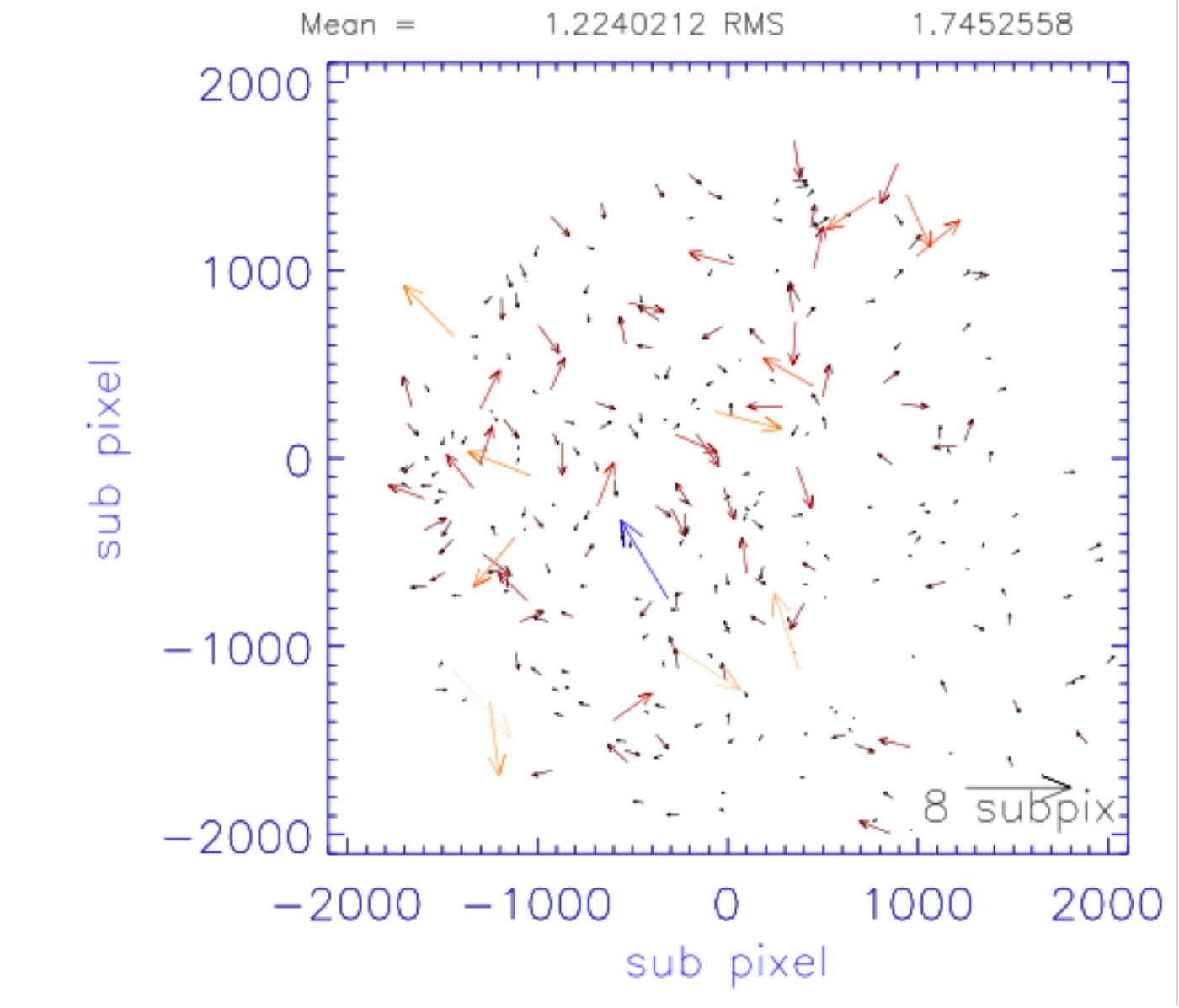}
\caption{Relative differences in positions  of the stars in the UVIT image taken with NUV (N279N) and the DSS-blue image are shown as vectors, where tail of the vector corresponds to position in 
the field. Positions and errors  (shown on the top) are shown in units of sub-pixel which is equal to ~ 0.41\arcsec.  
 More than 40\% of stars in the UVIT images could be matched with the selected bright stars ( a total of 1025 over 45\arcmin$\times$45\arcmin) of DSS within a distance of 8 sub-pixels (3.3\arcsec). The average deviation is 1.2 sub-pixels (0.7\arcsec), but some of the stars show large deviations. On checking with the DSS image, it is found that each and every match with deviation $>$ 4 sub-pixels (1.7\arcsec) corresponds to overlapping stars in DSS.
}
\end{center}
\end{figure*}

\section{Calibrations to be done in the future}

 The hardware of UVIT and its interfaces with the Spacecraft are designed to give an absolute timing accuracy of $<$ 5 ms 
for the images. However, due to some difficulties in data analysis the absolute timing cannot yet be obtained. In the 
future attempts would be made to get accurate time. At present only short term relative accuracy of few ms per 1000 sec 
is possible with internal clock of UVIT.  Further, more calibrations would be
carried out to obtain: a) flat-field data for all the filters with a good coverage extending to edges of the field, b) astrometric data for improving the corrections for distortion.

\section{Summary}
The calibrations of in-orbit performance of UVIT have been presented. The overall performance of the instrument has been 
consistent with the calibrations done on the ground. In particular, the sensitivity in FUV and NUV is found to be within 
80\% to 90\% of the expected, the spatial resolution in FUV and NUV exceeds the expectation, and relative astrometric 
accuracy over the field is about 0.5\arcsec (rms). Table 4, the effective area curves for all the filters and gratings, and dispersions and resolutions for the gratings will be made available in the UVIT website, http://uvit.iiap.res.in/
 The calibrations for absolute timing accuracy are yet to be done, 
and more data are required to fully characterise flat-field variations for all the filters.        

\begin{table*}[h]
\begin{center}
\caption{
The percentage of the counts as a function of radial distance from the center. The radius is in pixel, where 1pix = 0.413\arcsec. 
}
\begin{tabular}{ccc}
\small
Radius (pix)  &Percentage Flux( FUV) & Percentage Flux (NUV)\\
\hline
1	&     13.46  & 14.98\\
2	  &    38.35 & 40.74\\
3	   &   57.27 & 58.28\\
4         &   68.82  & 68.14 \\
5	   &   75.62  & 73.81\\
6	   &   80.09  & 77.82 \\
7	   &   83.29  &80.85\\
8	   &   85.70  &83.63 \\
9	    &  87.71  &83.63 \\
10         &  89.45  &88.21 \\
11	    &   90.85 &90.29 \\
12         &   92.12 &92.10 \\
13        &    93.27 &93.39 \\
14         &   94.26 &94.40\\
15          &  95.07  &95.22 \\
16          &  95.77 &95.94 \\
17	      &  96.44 & 96.69 \\
18	      &  97.07 &97.30 \\
19	      &  97.59 & 97.77\\
20	      &  98.03  &98.26 \\
22	      &  98.73  &99.04 \\
24           & 99.22 & 99.55\\
26          &  99.65 & 99.90\\
27          &   99.78 & 99.95 \\
\hline
\end{tabular}
\end{center}
\end{table*}

\acknowledgements
We thank the referee for helpful suggestions.
UVIT project is a result of collaboration between IIA, Bengaluru, IUCAA, Pune, TIFR, Mumbai, several centres of ISRO, and CSA. Several groups from ISAC (ISRO), Bengaluru, and IISU (ISRO), Trivandrum have contributed to the design, fabrication, and testing of the payload. The Mission Group (ISAC) and ISTRSAC (ISAC) of ISRO have provided support in making the observations, and reception and initial processing of the data.  We gratefully thank all the members of various teams for providing support to the project from the early stages of design to launch and observations in the orbit.

\end{document}